\documentclass[aps,twocolumn]{revtex4}

\usepackage{graphicx}
\usepackage{color}
\usepackage{subfig}

\begin{document}

\title{Separating Gravitational Wave Signals from Instrument Artifacts}
\author{\surname {Tyson} B. Littenberg}\affiliation{Maryland Center for Fundamental Physics, Department of Physics, University of Maryland, College Park, MD  20742}\affiliation{Gravitational Astrophysics Laboratory, NASA Goddard Spaceflight Center, 8800 Greenbelt Rd., Greenbelt, MD  20771}
\author{\surname{Neil} J. Cornish}\affiliation{Department of Physics, 
  Montana State University, Bozeman, MT 59717}
\date{\today}

\begin{abstract}
Central to the gravitational wave detection problem is the challenge of separating features in
the data produced by astrophysical sources from features produced by the detector.
Matched filtering provides an optimal solution for Gaussian noise, but in practice, transient
noise excursions or ``glitches'' complicate the analysis.  Detector diagnostics and coincidence
tests can be used to veto many glitches which may otherwise be misinterpreted as gravitational wave
signals.  The glitches that remain can lead to long tails in the matched filter search statistics and drive up the detection threshold. Here we describe a Bayesian approach that incorporates
a more realistic model for the instrument noise allowing for fluctuating noise levels that
vary independently across frequency bands, and deterministic ``glitch fitting'' using wavelets
as ``glitch templates'', the number of which is determined by a trans-dimensional
Markov chain Monte Carlo algorithm.  We demonstrate the method's effectiveness on simulated data
containing low amplitude gravitational wave signals from inspiraling binary black hole systems,
and simulated non-stationary and non-Gaussian noise comprised of a Gaussian component with the
standard LIGO/Virgo spectrum, and injected glitches of various amplitude, prevalence, and variety.
Glitch fitting allows us to detect significantly weaker signals than standard techniques.
\end{abstract}

\maketitle

\section{Introduction}
\label{intro}
To take full advantage of gravitational wave detectors we must confidently distinguish weak gravitational
wave (GW) signals from detector noise.  This challenge, dubbed the ``detection problem,'' warrants
significant attention as we prepare for the first direct measurement of gravitational waves within
the next decade. In their current operational configuration, the Laser Interferometer Gravitational
wave Observatory (LIGO)~\cite{LSC} and Virgo~\cite{Virgo}, encounter frequent large-amplitude transient
noise events, or ``glitches'', while detectable GW signals are forecast to be rare and weak. The LIGO-Virgo
collaboration has developed several practical techniques for addressing the detection problem in this
context (see {\it e.g.} Refs.~\cite{Abbott:2009zi,Abbott:2009tt}), but there is considerable
room for improvement.

We propose a new approach to the detection problem that works by simultaneously modeling the instrument
noise and the gravitational wave signals. It can be argued that the existing search algorithms
incorporate {\em implicit} noise modeling as the search statistics are tuned using
time slides of the data (the time shifts destroy the coherence of the gravitational wave signals
while preserving the noise properties in a statistical sense). In our Bayesian approach to the detection
problem we must use an {\em explicit} noise model, which in turn defines the likelihood function.
We wish to stress that the issue is not whether a Bayesian or frequentist approach is superior, but
rather that the ``tuning'' of the search must proceed in a different fashion. In the frequentist approach,
one or more statistics are chosen that (hopefully) indicate whether a gravitational wave signal is
present in the data. Time slides and signal injections are then used to produce estimates of the false
alarm and false dismissal probabilities for the statistics. The choice of statistics can be refined during
this procedure in an effort to minimize the false dismissals and false alarms. In a Bayesian approach
the analysis is fully determined by the choice of likelihood function and prior. Once these have been
chosen the analysis is purely mechanical, there are no thresholds to set or statistics to tune.
As Laplace put it ``the theory of probabilities is basically just common sense reduced to
calculus''~\cite{Laplace:1814}.
Though the well defined probability calculus of Bayesian analysis is appealing, the output is only as
good as the input. Bayesian analyses of LIGO-Virgo data that assume a Gaussian
likelihood function~\cite{vanderSluys:2007st,vanderSluys:2008qx,vanderSluys:2009bf} should be treated
with caution.

Rather than making guesses about the noise model, and hence the form of the likelihood function,
Bayesian inference can be used to determine the noise model from the data~\cite{Allen:2002jw}. To this
end, we introduce several parameterized models for the noise, and use the data to jointly estimate
the noise and signal parameters. Bayesian model selection is applied to alternative parameterizations
of the noise in an effort to find the most parsimonious representation. The process of designing these
parameterized likelihood models is similar to the process of designing statistics for a frequentist
analysis. The main difference is that the ``tuning'' of the likelihood models occurs mechanically,
without fear of operator induced biases.

We consider a variety of noise models, and in tests performed on simulated LIGO/Virgo noise, we find
that a model that combines a description of Gaussian noise with a time varying power spectrum and
a wavelet model that is able to fit semi-coherent glitches best represents the data. While we were careful
to ensure that the models used to generate the simulated data did not correspond exactly to the models
being tested, it is certainly no surprise that the model favored by the analysis is the one that is
most similar to the model used to simulate the data. Indeed, the main motivation for working with
simulated data is to have full control of the experiment - if the analysis had not selected the
likelihood model that is closest to the injected noise model we would know that something was wrong.
The next step, which we are actively pursuing, is to repeat our analysis using real LIGO/Virgo data.

We are not alone in promoting the idea that better noise modeling is key to the gravitational wave detection
problem. Allen {\it et al.}~\cite{Allen:2002jw,Allen:2001ay} have argued that parameterized noise models are
needed to derive robust search statistics. Clark {\it et al.}~\cite{Clark:2007xw} included a model
for Sine-Gaussian instrument glitches as an alternative hypothesis when computing Bayesian odds ratios
for gravitational wave signals from pulsar glitches.  Clark {\it et al.} also suggested that it would be
valuable to have a classification of instrument glitches that could be used to construct better
models for the instrument noise. Similar ``glitch hypotheses'' were considered
by Veitch and Vecchio~\cite{Veitch:2009hd} in a study of Bayesian model selection applied to the search
for black hole inspiral signals. The possibility of subtracting instrument glitches from the data
using a physical model of the detector has been investigated by
Ajith {\it et al.}~\cite{Ajith:2007hg,Ajith:2009}. Principe and Pinto~\cite{Principe:2008bz} have
introduced a physically motivated model for the glitch contribution to the instrument noise, and have
used this model to derive what they refer to as a ``locally optimum network detection
statistic''~\cite{Principe:2009zz}. In their approach the glitches are treated in a statistical sense,
while our approach directly models the glitches present in the data. The possibility of
directly deriving likelihood functions from the data in a Bayesian setting has previously been
considered by Cannon~\cite{Cannon:2008zz}. In Cannon's approach the data is first reduced to an n-tuple
of quantities, such as the parameters produced by a matched filter search, then using time slides and
signal injections the likelihood distributions for the signal and noise hypotheses are directly estimated
from the data. These are then used to estimate the posterior probability that a measured set of
parameters corresponds to a gravitational wave event.

We develop our approach as follows: In section~\ref{Like} we discuss the connection between noise models
and likelihood functions, and in section~\ref{wavelet} we describe the wavelet representation that we
use to define our noise models. Section~\ref{MCMC} provides a brief review of the Markov Chain Monte Carlo
algorithm that we use to carry out the calculus of Bayesian inference. In section~\ref{noise} we introduce
our noise models and describe the simulated data sets. Bayesian model selection is then applied to
identify the most effective noise model. Section~\ref{waveform} defines the black hole inspiral waveforms
that we inject, and section~\ref{detection} describes the search phase of our analysis. Results are
presented in section~\ref{results} describing a simultaneous characterization of gravitational wave
signals and instrument noise for two simulated data sets. We close with a discussion of our plans for
future work in section~\ref{discussion}.

\section{Likelihood Models}
\label{Like}
For Gaussian noise there is a well known optimal solution to the detection problem based on
Wiener matched filtering~\cite{Thorne:1987}, which can be described in terms of a detection
statistic $\rho$ - the matched filter signal-to-noise ratio, or equivalently, the log
likelihood $\log{\cal L} = -\chi^2/2$. Suppose the data $s$ produced by a
gravitational wave detector were
the superposition of instrument noise $n$ and the detector's response to a passing
gravitational wave $h^s$.  Then the signal-to-noise for a template $h$ is defined:
\begin{equation}
\rho(h) = \frac{(s \vert h)}{(h \vert h)^{1/2}} \, ,
\end{equation}
and the chi-squared residual by
\begin{equation}
\chi^2(h) = (s -h \vert s - h) \, .
\end{equation}
The expectation value of $\rho$ is maximized, and $\chi^2$ is minimized, when the template matches
the signal, $h=h^s$. Here we have used the standard noise weighted inner product for Gaussian
noise with one-sided noise spectral density $S_n(f)$:
\begin{equation}\label{fnwip}
(a\vert b) = 2 \int \frac{a b^* + a^* b}{S_n(f)} \, df \, .
\end{equation}

The gravitational wave detection problem is usually described in the frequentist language of
false alarms and false dismissals based on Monte Carlo studies of a suitably chosen detection
statistic. According to the Neyman-Pearson theorem, the quantity $\rho$ provides an optimal
statistic for stationary, Gaussian noise as it yields the maximum detection probability for
a given false alarm probability. While there are no similar proofs of optimality for the
$\rho$ statistic when applied to the non-stationary and non-Gaussian noise encountered in
the LIGO/Virgo detectors, the matched filter SNR or closely related quantities are adopted
as search statistics that are then calibrated using Monte Carlo studies of signal injections and
scrambled data. Sections of data that are identified as being corrupted by excessive noise
or transient instrumental artifacts (glitches) are vetoed prior to the tuning of the search
statistic. The development of these vetoes is a complex topic, but the basic idea is to either
look for correlations between glitches in the gravitational wave channel and the thousands of
diagnostic channels, or to look for statistical patterns. A more detailed description of the
vetoing procedures and data quality assessment can be found in Ref.~\cite{Abbott:2009zi}.

An alternative to tuning a statistic chosen for its performance with Gaussian noise is to
look for new statistics that can be shown to be near optimal for relevant forms of non-Gaussian
noise~\cite{Allen:2001ay,Allen:2002jw}. This approach is similar in spirit to what we are proposing.

It is interesting to contrast the frequentist and Bayesian approaches to the detection problem.
In the Bayesian approach we simply compute the posterior distribution function $p(h\vert s,\mathcal{M})$ for
the waveforms $h$ of model hypothesis $\mathcal{M}$, and the associated model evidence $p(s\vert \mathcal{M})$. These are
related by Bayes' theorem to the likelihood $p(s\vert h,\mathcal{M})$ and the prior $p(h\vert \mathcal{M})$ by
\begin{equation}~\label{Intro: bayes}
p(h \vert s, \mathcal{M}) = \frac{p(h\vert \mathcal{M}) p(s\vert h,\mathcal{M})}{p(s\vert \mathcal{M})} \, ,
\end{equation}
and
\begin{equation}~\label{Intro: evidence}
p(s\vert \mathcal{M}) = \int p(h\vert \mathcal{M}) p(s\vert h,\mathcal{M}) dh \, .
\end{equation}
Everything we want to know about the waveform model $\mathcal{M}$ is contained in the posterior distribution,
while the evidence allows comparisons to be made between alternative models ({\it e.g.} is a GW
signal present or not). It is important to emphasize that the Bayesian approach is entirely mechanical, there
are no statistics to tune or thresholds to set - you simply go ahead and calculate confidence
intervals for the waveform parameters and odds ratios for competing hypotheses. On the other
hand, the output of this mechanical process is only as good as the inputs - if the waveform model
or the likelihood function is flawed, then the conclusion will also be flawed. We
will assume that the waveform model is accurate (see Ref.~\cite{Cutler:2007mi} for a discussion
of the biases introduced by inaccurate waveform models), and concentrate our attention on the
likelihood function. 

The data collected by one or more gravitational wave detectors can be
represented by a discrete collection of samples $s_i$. These may represent the strain sampled
at certain times, Fourier amplitudes, wavelet amplitudes {\it etc}. In the absence of a
gravitational wave signal, $s_i=n_i$, and we are looking at samples drawn from the instrument
noise distribution $p_N(n_i)$. When a signal $h_i$ is present we have $n_i=s_i-h_i$, and assuming
that the noise and the
signal are uncorrelated, the likelihood of observing $s_i$ given the waveform $h_i$
is simply~\cite{Finn:1992wt}
\begin{equation}
p(s_i \vert h_i, \mathcal{M}) = p_N(s_i-h_i) \, ,
\end{equation}
while the likelihood of observing the full data set $\{s_1,s_2,\dots, s_N\}$ is the
joint probability
\begin{equation}
p(s \vert h, \mathcal{M}) = p_N(s_1-h_1, s_2-h_2, \dots, s_N-h_N) \, .
\end{equation}
If the noise in each sample is independent, then the joint distribution is the
product of the individual distributions:
\begin{equation}\label{Intro: likelihood}
p(s \vert h, \mathcal{M}) = \prod_{i=1}^N p_N(s_i-h_i) \, .
\end{equation}
The key point is that the likelihood function is nothing other than the
probability distribution that describes the instrument noise. And therein lies the problem:
The description of the likelihood function, and hence the veracity of the Bayesian approach,
is only as good as our understanding of the instrument noise. A solution to this problem
is to introduce a model for the noise~\cite{Allen:2002jw}, and to use the
data to jointly estimate the noise and signal parameters.  We have taken this approach
to a limited extent in previous analyses where we treated the noise as Gaussian, but with
its variance to be determined from the data~\cite{Cornish:2007if, Littenberg:2009bm, Adams:2010vc}.
Here we extend the scope of the noise modeling to consider non-Gaussian tails and
inter-sample noise correlations of the type caused by semi-coherent ``glitches''.

\begin{figure}[htbp]
   \includegraphics[angle=0, width=0.5\textwidth]{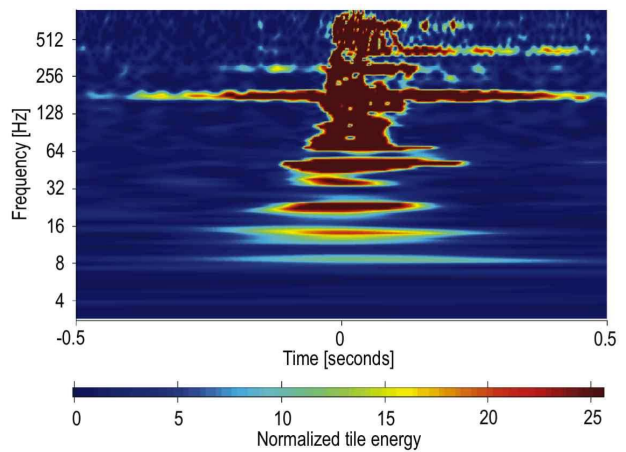} % requires the graphicx package
   \caption{A time-frequency scaleogram showing a loud glitch in the output of
one of the LIGO detectors (LIGO document number LIGO-G070807-00-Z).}
   \label{glitch scaleogram}
\end{figure}

Recall that for white Gaussian noise the samples $n_i$ will be independent in time and
in frequency, while for colored Gaussian noise the samples will be independent in frequency
and correlated in time. To avoid the complications of having to account for these correlations
with a joint probability distribution, it is standard practice to compute the likelihood in
the frequency domain, where the likelihood factors into a simple product~\cite{Finn:1992wt}.
However, as is clear from Figure~\ref{glitch scaleogram}, the glitches seen in the LIGO/Virgo data
are semi-coherent in time and frequency - that is, if there is excess noise in one time
frequency pixel, then there is a high probability that there will be excess noise in a
neighboring time-frequency pixel. These glitches not only produce large non-Gaussian tails,
they also introduce strong correlations in the noise model. While there are many ways to try
to model this behavior, we will show that treating the glitches as coherent instrumental artifacts
provides an effective solution.

\section{Data Analysis in the Wavelet Domain}
\label{wavelet}
The first step towards constructing a realistic noise model is to abandon the familiar Fourier-domain approach to signal processing in favor of a time-frequency decomposition in the wavelet domain.

Time-frequency methods are employed in a variety of GW search algorithms for LIGO/Virgo data, particularly in searches for gravitational wave ``bursts'' -- transient gravitational wave signals for which no templates are available -- such as the coherent Wave Burst (cWB) algorithm~\cite{Klimenko:2008fu}.  
\begin{figure*}[htbp]
\subfloat[time]{\includegraphics[width=0.48\textwidth]{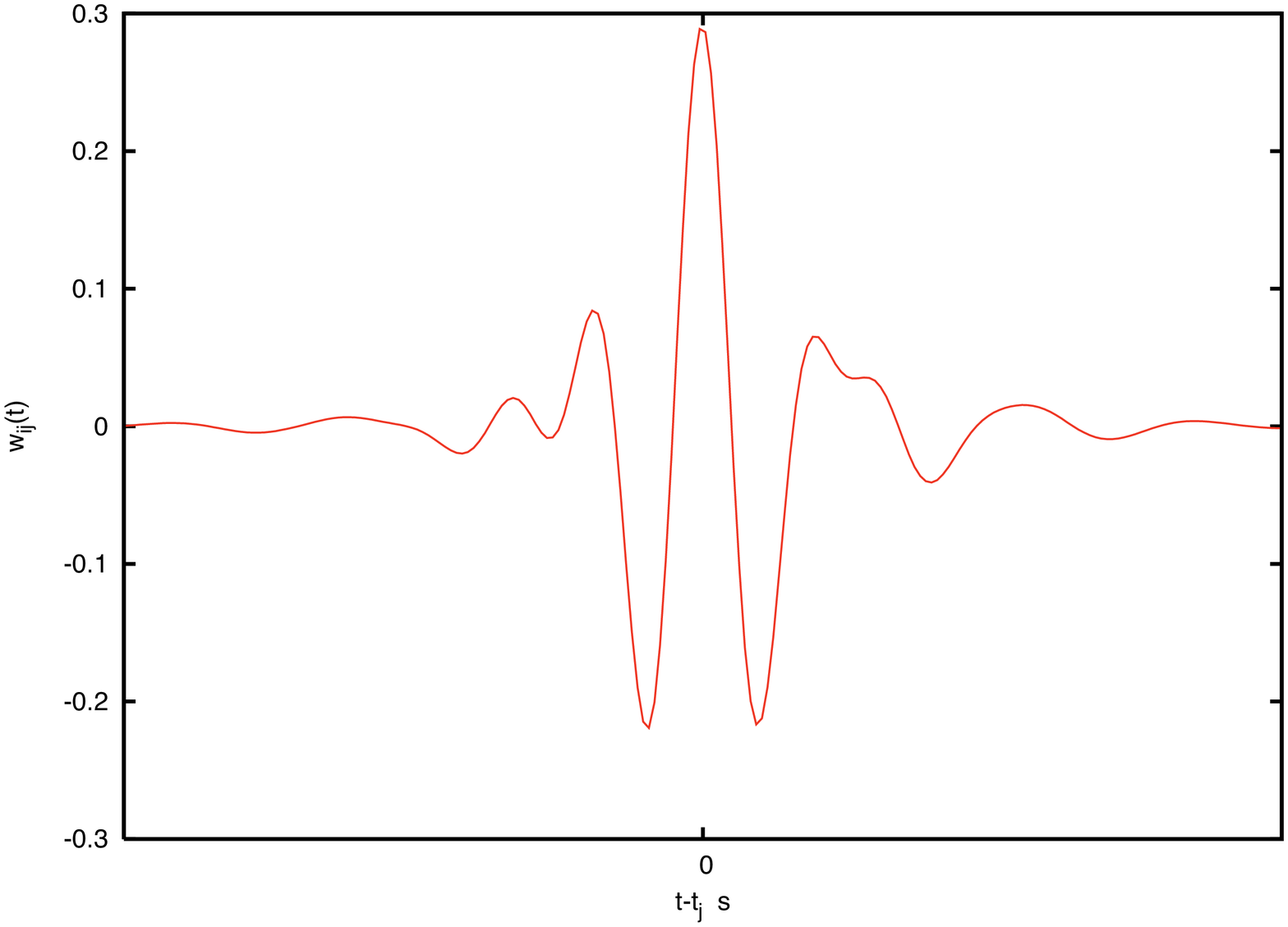}}
\subfloat[frequency]{\includegraphics[width=0.48\textwidth]{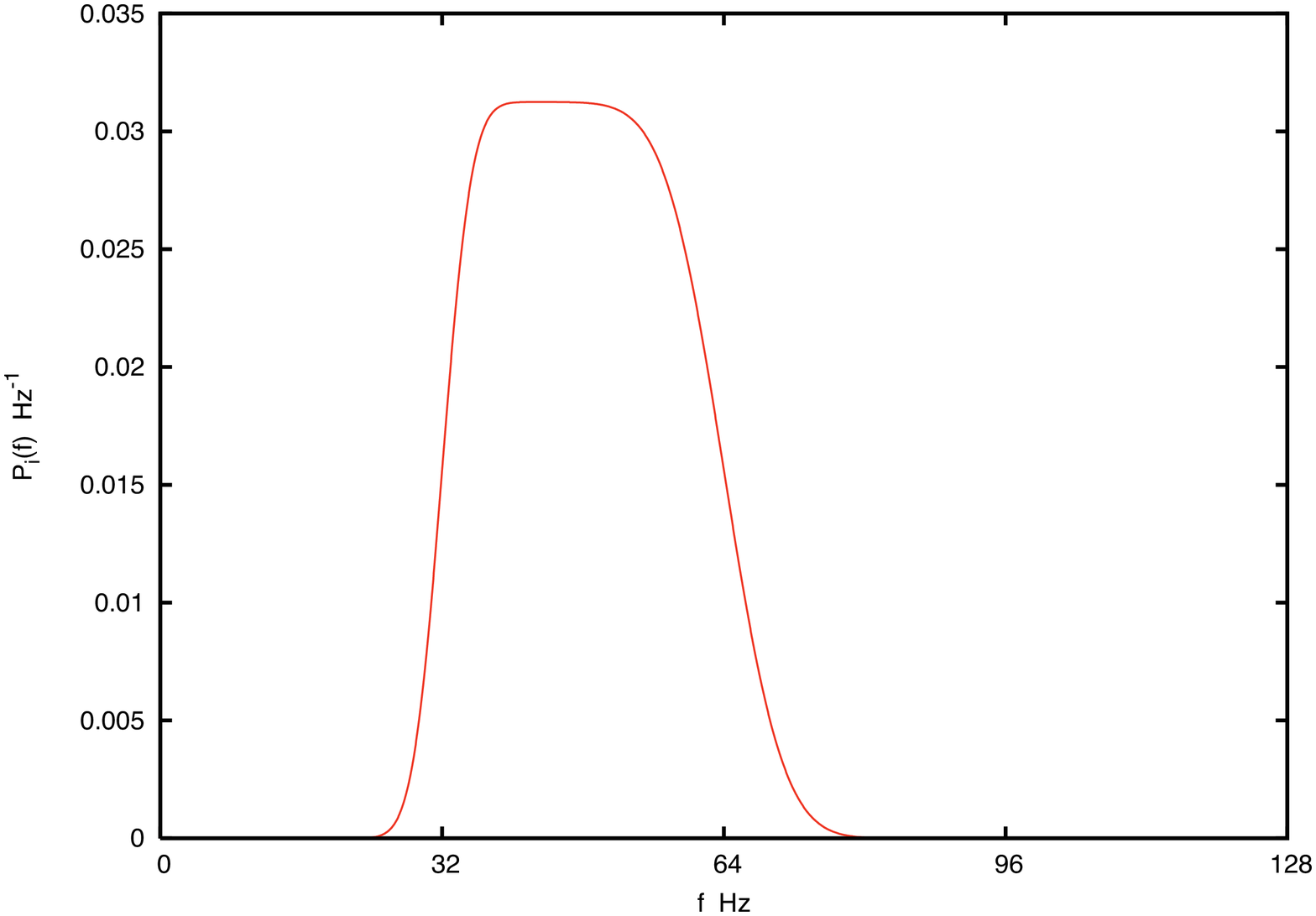}}
\caption{The Meyer wavelet basis function for frequency layer $i=9$ and signal duration of 16 seconds.  a) Time domain $\psi_{ij}(t)$ for arbitrary time index $j$.  b) Fourier power spectrum of $\tilde{\psi}_{ij}(f)$.  Notice how, for $i=9$ and $T=16$ s, this wavelet acts as a bandpass filter for frequencies $f\sim[32,64]$ Hz.}
\label{BH: wavelets}
\end{figure*}

Individual wavelet functions $\psi_{ij}$ are compact in both frequency and time, and can be used to form an orthogonal basis~\cite{NRC}.  Each element of the basis is a scaled, time-shifted version of the ``mother wavelet'' $\psi$ of which a large variety exist.  We have chosen the Meyer wavelet to form our basis functions as is used in cWB.  Fig.~\ref{BH: wavelets} depicts an example of the Meyer wavelet in the time domain, as well as its power spectrum.  The Fourier power of a single wavelet function is perhaps most illuminating.  Each wavelet basis function acts as a bandpass filter, selecting for a specific range of frequencies, during a specific time interval of the function being decomposed.  

We use the discrete wavelet transform (DWT) of a time series $s(t)$ to determine the wavelet coefficients $w_{ij}$.  Each $w_{ij}$ represents the amplitude of a pixel in the time-frequency plane with volume $\Delta t \Delta f = 1/2$.  The indices $i$ and $j$ correspond to the frequency and time, respectively, of the wavelet coefficient. 

\begin{figure}[!h]
%\begin{center}
\includegraphics[width=0.5\textwidth]{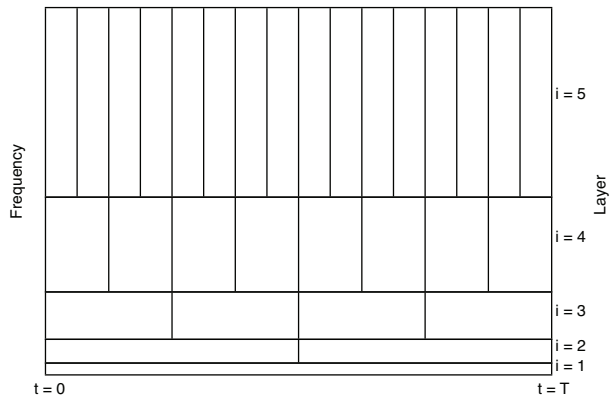}
\caption{A cartoon depicting the tiling of the time-frequency plane by a discrete wavelet transform (DWT).  Each pixel has time-frequency volume $\Delta t \Delta f = 1/2$. }
\label{BH: wavelet tile}
%\end{center}
\end{figure}

Unique to the DWT is the way the pixels ``tile'' the time-frequency plane:  Low frequency wavelets have long durations and narrow frequency response.  As we move to higher frequency the bandwidth of the wavelet increases while the duration shortens (maintaining $\Delta f \Delta t=1/2$).  A cartoon representation of this time-frequency tiling can be found in Fig.~\ref{BH: wavelet tile}.  Each frequency layer, denoted by the index $i$, contains $2^i$ divisions in time and, for signal duration of $T$, is $2^{i-1}/T$ wide in frequency.

For stationary, Gaussian noise, the wavelet coefficients are sufficiently uncorrelated to treat the
wavelet-domain noise-correlation matrix as diagonal.  In this instance the expectation value of the
noise power for the wavelet pixel with indices $i$ and $j$, $S_{ij}$, can be computed by filtering
the noise power spectral density $S_n(f)$ with $P_w^{ij}(f)  = |\tilde{\psi}_{ij}(f)|^2$:
\begin{equation}
S_{ij} = \int_0^{\infty} P_w^{ij}(f) S_n(f) df.
\end{equation} 
The power spectrum for any wavelet in layer $i$, as well as the power spectral density at times $j$
are the same for stationary, Gaussian noise. Including the time index will prove useful
when we allow for noise varying with time.

We are now able to define a diagonal noise-weighted inner product as 
\begin{equation}\label{nwip}
(a|b) \equiv \sum_k\sum_{i,j}\frac{a^k_{ij}b^k_{ij}}{S_{ij}^k}.
\end{equation}
where the index $k$ represents an individual interferometer, or IFO, in the network. For
Gaussian noise, the inner products computed in the wavelet domain using (\ref{nwip})
are almost identical to the the inner products computed in the Fourier domain using (\ref{fnwip}).

\section{Markov Chain Monte Carlo}
\label{MCMC}
Bayesian methods have lagged behind their frequentist brethren because of the burden associated with evaluating Eqs.~(\ref{Intro: bayes}) and~(\ref{Intro: evidence}).  Recently, powerful
techniques for overcoming these computational hurdles (e.g., Markov Chain Monte Carlo (MCMC)
methods~\cite{Metropolis:1953, Hastings:1970}, Nested Sampling~\cite{Skilling:NS}, etc.)
have become progressively more efficient and computer power has increased,
allowing us to employ these resources on
interesting GW detection problems~\cite{Christensen:1998gf,Christensen:2001cr}.  

Our method of choice for computing the posterior distribution function are the MCMC algorithms.  These techniques have shown prowess in solving parameter estimation problems, while variants of these methods have also proven effective as search algorithms~\cite{Cornish:2005qw, Crowder:2006eu, Cornish:2007jv, Brown:2007se, Gair:2008, Robinson:2008fb, Trias:2008dc,Key:2008}.  Here we briefly describe the basis of the algorithms, and leave specific details of our implementation to later sections.

MCMCs provide samples from the (previously unknown) target posterior distribution function for the parameters of some model $\mathcal{M}$.  This is accomplished by first adopting a position in parameter space $\vec{\theta}_x$ as the first ``link" in the chain and evaluating that position's likelihood $p(s|\vec{\theta}_x,\mathcal{M})$ and prior probability $p(\vec{\theta}_x|\mathcal{M})$.  Next, we suggest a trial position, $\vec{\theta}_y$,  from the proposal distribution $q(\vec{\theta}_y|\vec{\theta}_x)$ read \emph{``the probability of suggesting a move to $\vec{\theta}_y$ given that the current
location is $\vec{\theta}_x$"}.  The new likelihood and prior probability are evaluated and $\vec{\theta}_y$ is adopted as the next link in the chain with probability ${\kappa} =\text{min}[1,H]$ where $H$ is the Hastings ratio
\begin{equation}
H_{\vec{\theta}_x \rightarrow \vec{\theta}_y}=\frac{ p(s|\vec{\theta}_y,\mathcal{M}) p(\vec{\theta}_y\vert \mathcal{M}) q(\vec{\theta}_x|\vec{\theta}_y) }{ p(s|\vec{\theta}_x,\mathcal{M}) p(\vec{\theta}_x \vert \mathcal{M}) q(\vec{\theta}_y|\vec{\theta}_x) }.
\end{equation}
This process of stochastically stepping through parameter space repeats until some convergence criteria are satisfied (See Ref.~\cite{Littenberg:2009bm} for a description of the convergence tests we
use).  Afterwards, the number of iterations spent in a particular region of parameter space, normalized by the total number of steps in the chain, yields the probability that the model parameters have values within that region.

The Hastings ratio is derived by mandating transitions from $\vec{\theta}_x$ to $\vec{\theta}_y$ satisfy the detailed balance condition which ensures the samples from the Markov chain are representative of the target PDF. 

The choice of $q(\vec{\theta}_y|\vec{\theta}_x)$, by construction, can not alter the recovered posterior distribution function. The proposal distribution does, however, dramatically affect the acceptance rate of trial locations in parameter space and, therefore, the number of iterations required to satisfactorily sample the joint PDF.

Markov chains are still prone to being ``trapped'' by local maxima of the target distribution for longer than the user would be willing to wait.  Some enhancement to the above prescription is often mandatory in order to ensure global sampling of the PDF.  One such method, parallel tempering~\cite{Swendsen:1986} , uses a set of chains running simultaneously, each at a higher ``temperature.''  The likelihood of a chain with inverse temperature $\beta=1/T$ is calculated by $p(s|\vec{\theta},\mathcal{M})^{\beta}$.  Increasing the temperature effectively smoothes the topography of the distribution being explored.  High temperature (low $\beta$) suppresses the differences in likelihood between the current and proposed model parameters, allowing the high temperature chains free exploration of the parameter space.  In the limit where the temperature goes to infinity ($\beta$ goes to zero) the chain will sample the prior distributions.

Parallel tempering promotes adequate mixing by allowing the chains of different temperature to exchange parameters, thus allowing solutions found by the hot chains to be communicated to the colder chains.  This sharing of solutions can be executed while maintaining detailed balance if exchanges are accepted using  
\begin{equation}
H_{i\leftrightarrow j}=\frac{p(s|\vec{\theta}_i, \mathcal{M}_i, \beta_{j})p(s|\vec{\theta}_{j},  \mathcal{M}_{j},\beta_{i})} {p(s|\vec{\theta}_i, \mathcal{M}_i, \beta_{i})p(s|\vec{\theta}_{j}, \mathcal{M}_{j}, \beta_{j})}
\end{equation}
as the Hasting's ratio in the transition probability for an exchange between chain $i$ and $j$.  Only points in the $\beta=1$ chain sample the target distribution and are therefore permitted to contribute to the chain from which the PDF is inferred.  However, by exchanging parameters with hotter chains the $\beta=1$ samples rapidly explore the full target distribution, including movement between different modes of the posterior which, for practical applications, can be impossible for a single chain to achieve.

While only the $\beta=1$ chain contributes to the PDF, the rest of the chains serve as more than just a convergence aid.  The average log-likelihood of each chain, integrated over $\beta = [0,1]$, is equivalent to the integral in (\ref{Intro: evidence})~\cite{Goggans:2004}, allowing us to perform parameter estimation and model selection studies using a single analysis ``pipeline.''   

It should be noted that simultaneously solving the parameter estimation and model selection problem is not a unique attribute of the parallel tempered  Markov chain Monte Carlo (PTMCMC) algorithm.   On the contrary, the MultiNest algorithm~\cite{Feroz:2007kg, Feroz:2009de} also serves as an ``all-in-one'' Bayesian analysis package that has been successfully employed in GW data analysis problems.  While we will exclusively use the PTMCMC approach in this paper we want to emphasize that these algorithms are merely tools used to mechanically perform the calculation
of (\ref{Intro: bayes}) and (\ref{Intro: evidence}), and while some tools may be better
suited for a particular problem than others, the conclusions drawn from the data should
be identical.  

\section{Modeling the Noise} 
\label{noise}
We now return to our goal of finding the best noise model (equivalently the best likelihood function) to be used when describing realistic GW data.  We will construct three different likelihood functions and compare their performance on simulated non-Gaussian noise using Bayesian model selection.  What follows is a brief summary of each noise model before a more detailed description and comparison:
\begin{itemize}
\item Model $N_0$:  Allow the noise level to vary as a function of both time and frequency.
\item Model $N_1$:  Use a likelihood function derived from a distribution which has non-Gaussian tails.
\item Model $G_1$:  In conjunction with $N_0$ or $N_1$, model the noise as two independent contributions -- a stochastic component drawn from $N_i$ plus coherent ``glitches.''
\end{itemize}

\subsection{$N_0$:  Fitting to the noise level}
The first order solution is to fit to the underlying ``DC'' noise level in the data.  This has been done in frequency domain analyses~\cite{Cornish:2007if, Littenberg:2009bm, Adams:2010vc} allowing the model enough flexibility to account for errors in the predicted $S_n(f)$.  In the wavelet basis this approach to noise modeling gains sensitivity to non-stationarities because temporal information is encoded in the noise ``spectrum.''  

For this model, parameterized by $\vec{\eta}$, we use the expected noise level in ``blocks'' of wavelet pixels containing a certain time-frequency volume (TFV) normalized by the theoretical noise level,
 \begin{equation}
 \eta_{ij}^k = \frac{S_{ij,\rm{measured}}^{k}}{S_{ij,\rm{theoretical}}^{k}}
 \end{equation}
with a single value of $\eta$ assigned to each TFV block: 
 \begin{equation}
 \eta_{ij_0}^k = \eta_{i,j_0+1}^k = ... = \eta_{i,j_0+\rm{TFV}-1}^k \, ,
 \end{equation}
where $j_0$ is the time index of the first wavelet coefficient in the block under consideration.
Because the noise level is allowed to vary, the normalization of the likelihood and
the noise weighted inner product in the chi-squared depend on the noise parameters:
\begin{equation}\label{likelihood2}
\ln{p(s|\vec{\eta},N_0)} = -\frac{1}{2}\left((r|r) + \sum_{k}^{\rm{IFO}}\sum_{i,j}\ln{\eta_{ij}^k}\right)
\end{equation}
with  $\eta_{ij}^k S_{ij}^k$ substituted for $S_{ij}^k$ in the inner product defined
by (\ref{nwip}).  The second term in (\ref{likelihood2}) comes from the normalization
of the likelihood. Note that for a noise-only model, the residual $r$ is simply the data $s$.

If we fit to the noise level in large time-frequency blocks we lose sensitivity to short-duration non-stationarities, while using small time-frequency blocks results in a model with a large number of parameters, all of which must be constrained.  In a model selection sense, the former will produce poor fits to the data (if non-stationarities are present) while the latter will carry a large ``Occam Factor'' penalty for the model's additional degrees of freedom.  Some tuning is required to choose the correct TFV for the noise blocks in order to optimize this approach.  While $N_0$ is a non-stationary noise model, it is unable to respond to short duration
impulses of noise unless it is equipped with an unreasonable number of free parameters.  

\subsection{$N_1$:  Non-Gaussian Tails}
This noise model, denoted by $N_1$, will continue to fit to the noise level in blocks of wavelet pixels as in $N_0$, while additionally employing the suggestion in Ref.~\cite{Allen:2002jw} to redefine the likelihood as a weighted sum of two normal distributions.  The majority of the weight will be allocated to the distribution with the same variance as a Gaussian model so that the bulk of noise samples are presumably ``drawn from'' the standard picture of the instrument noise.  However, some non-zero contribution to the likelihood comes from a significantly wider distribution.  This broadens the tails of the noise distribution without greatly altering it's core.  In other words, the majority of samples drawn from this distribution will ``look'' as though they come from the ideal noise distribution, however the frequency of ``outliers'' is greatly increased.  

Symbolically, the probability of measuring $n_i$ in a single bin of the data can be calculated as
\begin{equation}\label{doublewide}
p(n_i) = \frac{1}{\sqrt{2\pi}}\left(\frac{\alpha}{\sigma_1}e^{-n_i^2/2\sigma_1^2} + \frac{1-\alpha}{\sigma_2}e^{-n_i^2/2\sigma_2^2}\right).
\end{equation}  
To demonstrate the effect of this distribution, suppose we choose $\sigma_2 = 3\sigma_1$ and $\alpha = 0.99$.  Then, for a single noise sample,
\begin{equation}
\frac{p(n=0|N_1)}{p(n=0|N_0)} = 0.993\, 
\end{equation}
and
\begin{equation}
\frac{p(n=10\sigma_1|N_1)}{p(n=10\sigma_1|N_0)} = 1.970\times10^{18} \, .
\end{equation}
We see that for samples near the mean, the difference between the two distributions is negligible,
while there is a large increase in the probability of large noise excursions under $N_1$.  

Following the discussion deriving Eq.~(\ref{Intro: likelihood}), the likelihood of measuring
the residual $r$ in the wavelet domain for $N_1$ becomes
\begin{equation}
p(s|\vec{\eta},N_1) = \prod_{k}\prod_{ij}p(r_{ij}^k).
\end{equation}

Using a superposition of two normal distributions is a standard method in Bayesian modeling of non-gaussian data~\cite{Sivia} but has not previously been implemented in gravitational wave signal processing.

\subsection{$G_1$:  Fitting for Glitches}
While both $N_0$ and $N_1$ possess many of the qualities which we desire in an accurate noise model, both assume that each noise sample in the data is independent (in the wavelet domain).  Instrument glitches, on the other hand, are coherent in time-frequency space.  A wavelet scaleogram representation, depicted in Figure~\ref{glitch scaleogram}, is a useful visualization of this feature, as the glitch can be seen ``lighting up'' a large cluster of pixels, while the remaining data samples are well described as being independently, randomly distributed.

We require some way of modeling the individual glitches, and a template bank of such events would undoubtedly be inefficient and incomplete.  Instead we can once more make use of the wavelet basis, and model the glitches as linear combinations of the basis functions.  Studies of LIGO/Virgo glitches using \emph{atoms} (e.g., sine-Gaussians) have shown that typical events can be decomposed with $\sim 10$ basis functions~\cite{Principe:2008bz, Principe:2009zz} so it is reasonable to anticipate a similarly small number of wavelets will be sufficient.  The parameters for the glitch fitting are the number of wavelets included in the fit, $n_G$, the indices identifying where in time-frequency space those wavelets (referred to as ``hot pixels'') are located, and their amplitudes $\vec{\gamma}$.  

It is unknown \emph{a priori} how many, or where in time-frequency space, these hot pixels will be needed.  Given this uncertainty we must automatically adjust the number and location of wavelet coefficients in the glitch fitting, a task best accomplished by the Reversible Jump Markov Chain Monte Carlo (RJMCMC)~\cite{Green:2003} approach.  This breed of MCMC is able to transition between models of differing dimension while satisfying detailed balance if the probability of a trans-dimensional move is computed using 
\begin{equation}\label{RJ Hastings}
H_{\mathcal{M}_i \rightarrow \mathcal{M}_j}=\frac{p(s|\vec{\theta}_j, \mathcal{M}_j)p(\vec{\theta}_{j},  \mathcal{M}_{j})q(\vec{\theta}_{i},\mathcal{M}_{i})}
	    {p(s|\vec{\theta}_i, \mathcal{M}_i)p(\vec{\theta}_{i},  \mathcal{M}_{i})q(\vec{\theta}_{j},\mathcal{M}_{j})} |J_{ij}|
\end{equation}
as the Hasting's ratio in the transition probability.  The Jacobian $|J_{ij}|$ accounts for any change in dimension between models $\mathcal{M}_i$ and $\mathcal{M}_j$.  The number of samples spent in each model, normalized by the total number of samples in the chain, is the posterior probability, or evidence, for that model (assuming adequate mixing, convergence, etc.).

In our glitch-fitting RJMCMC, trans-dimensional moves propose to either remove a single pixel from the glitch model (setting a wavelet coefficient to zero), or include a new pixel which was previously not a part of the glitch model (selecting a wavelet coefficient which is zero, and assigning it some amplitude).  

Because the number and location of hot pixels is variable, model $G_1$ is more aptly described as a ``meta-model,'' while a particular number and configuration of hot pixels represents a discrete model for the noise.  We can denote an individual glitch model by it's number of hot pixels, $n_G$.   For model $n_G$ and data containing $N$ samples, there are ``$N$ choose $n_G$''~= $K_n$ unique combinations of hot pixels.  We want each configuration to have equal prior probability, so the joint prior probability for each noise model configuration is
\begin{equation}
p(n_G|G_1) = 1/K_n = \frac{n_G!(N-n_G)!}{N!}.
\end{equation}
Meanwhile, we choose as the prior on the pixel amplitudes $p(\gamma_{ij}|n_G,G_1)=N[0,100S_{ij}]$.

RJMCMCs are often difficult to work with because of inefficient mixing between dimensions.  The key to overcoming this obstruction is to construct trans-dimensional proposal distributions which closely match the target distribution.  We do so by taking advantage of a glitch's tendency to excite a cluster of pixels in time-frequency space.  The trans-dimensional proposal distributions thus favor the inclusion of new pixels in the glitch model which are adjacent to the already existing clusters.  We do so by assigning each unoccupied pixel a probability of being selected for inclusion in the glitch model which is weighted by the number of neighboring pixels already ``lit up.''  Meanwhile, the trans-dimensional proposal distribution for the glitch amplitudes is identical to that of the prior on the amplitudes, thus canceling in the Hasting's ratio.

\subsection{Simulating Non-Gaussian Noise}
To test our approaches to noise modeling, before applying these techniques to data collected by the LIGO-Virgo network, we will analyze simulated data containing Gaussian noise with injected glitches.  We use a variety of glitch injections to ensure that any success is not a result of tuning the methods to a particular noise realization.  We begin by generating Fourier-domain noise for a 16 second data segment using the initial LIGO/Virgo design sensitivities.  We can inject glitches of two different type into the data.  One which we will refer to as the ``gaussian-gaussian'' type, is a burst of gaussian noise enveloped by a gaussian profile in time.  The second are the glitch atoms described in Ref.~\cite{Principe:2008bz}.  

Glitches are either placed in the data deliberately, or randomly chosen such that the glitches are distributed uniformly in the time-frequency plane, with average durations of $~10$ ``cycles.''  Each generated glitch $g_0$, is rescaled to achieve a desired ``SNR'' =~$\sqrt{\left(g|g\right)}$.  The glitch SNR, $x$, is either chosen by hand, or drawn from a double-power law distribution 
\begin{equation}\label{glitch distribution}
p(x) = C\left(x^{-4} + x_i^{-5/2}x^{-3/2}\right),\ \ x > x_0
\end{equation}
calibrated to the Black Hole ring-down trigger rate in a segment of ``typical'' LIGO S5 data.  $C$ is a normalization constant and $x_i\sim10$.  The total number of injections was also determined by the ring-down trigger rates.  This distribution produces glitches which are predominantly buried by the Gaussian component of the noise, with only a rare ($\sim 1$ in 5) noise realization containing an event visually above the Gaussian noise.  By injecting glitches from this distribution we are ruining the stationarity and Gaussianity of the noise without using our models for the noise to produce the samples.   

\subsection{Comparison of Noise Models}
To compare the relative performance of our new approaches to noise modeling, we analyzed three different signal realizations, $s_1$: Gaussian noise; $s_2$: Gaussian noise plus a single, loud (SNR $\sim100$) glitch; and $s_3$: Gaussian noise plus 300 glitches distributed across the network with SNRs drawn from
(\ref{glitch distribution}). We use the same Gaussian noise realization in each simulation.  

\begin{figure*}[htbp]
\subfloat[$s_1$]{\includegraphics[height=0.45\textwidth, angle = 270]{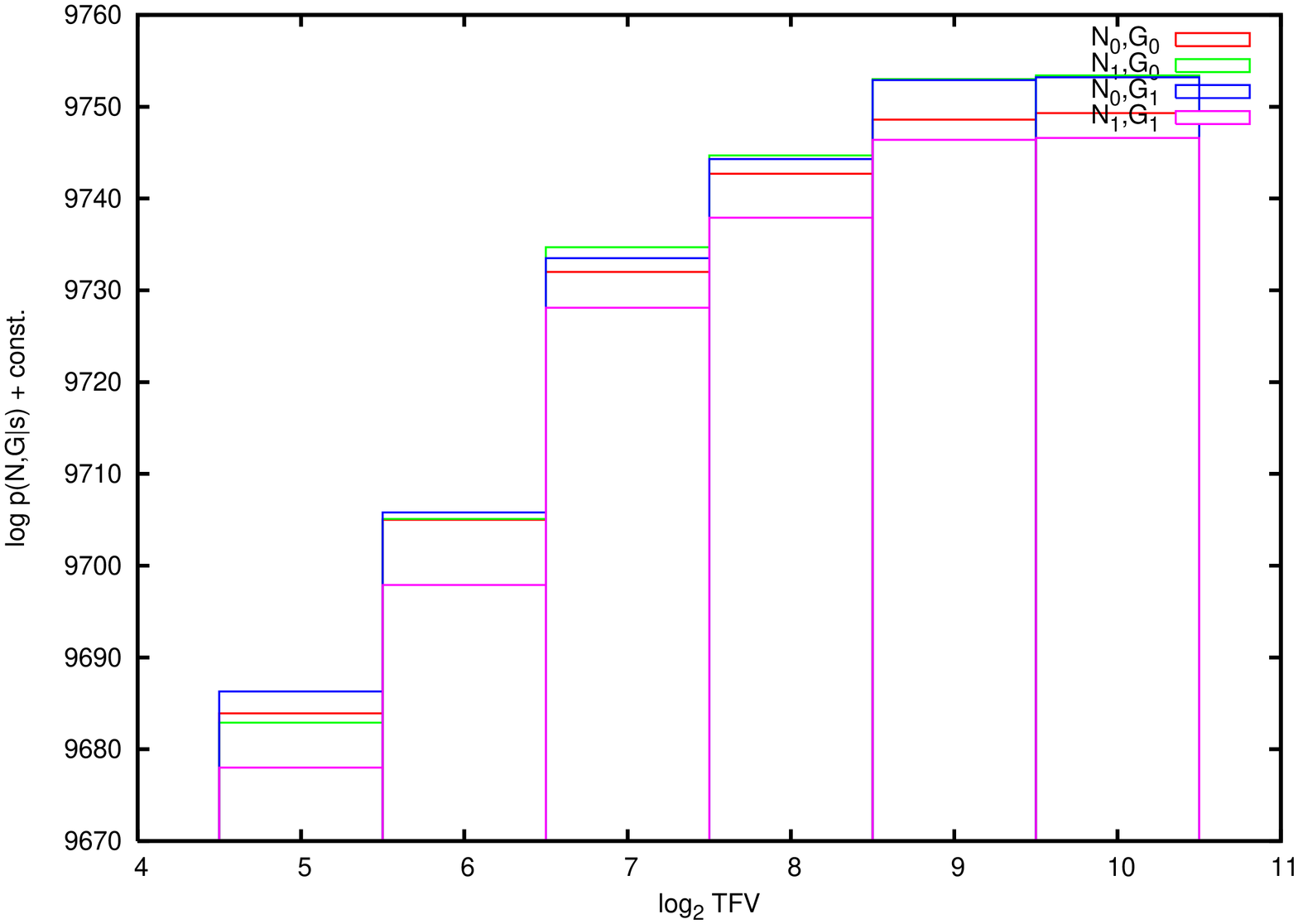}}
\subfloat[$s_2$]{\includegraphics[height=0.45\textwidth, angle = 270]{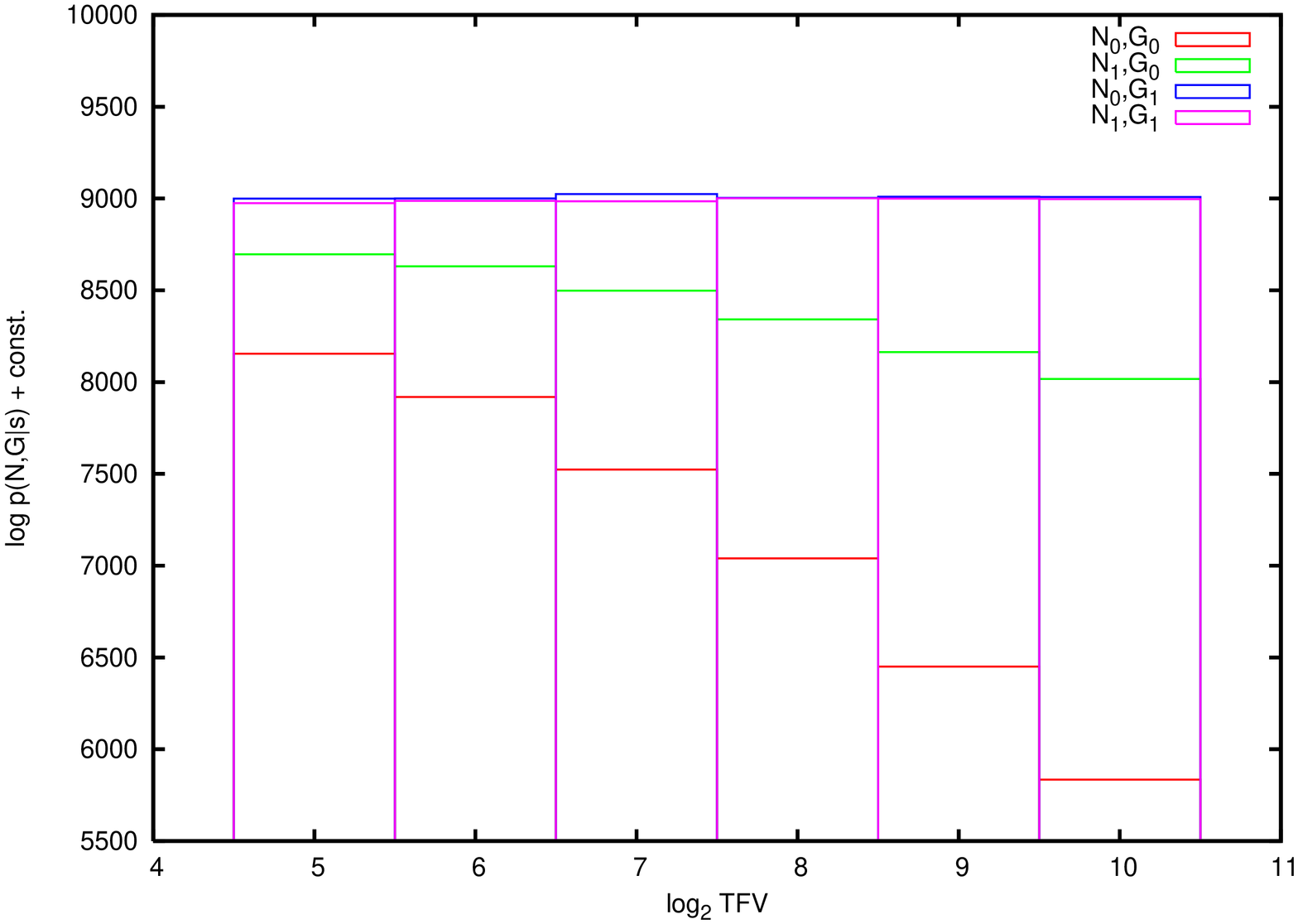}}\\
\subfloat[$s_3$]{\includegraphics[height=0.45\textwidth, angle = 270]{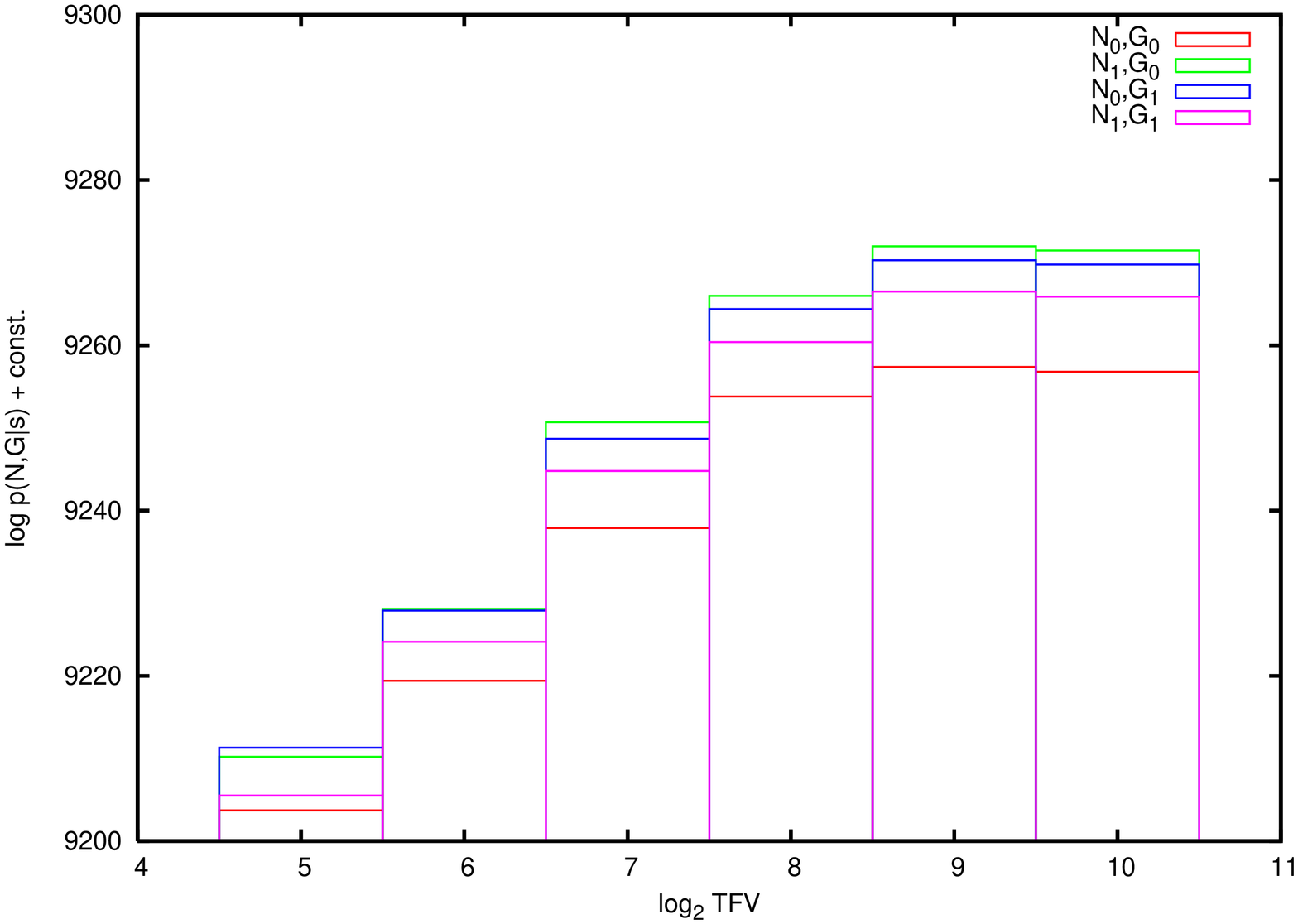}}
\subfloat[$p(m|N_i,G_1)$ for $s_3$]{\includegraphics[height=0.45\textwidth, angle = 270]{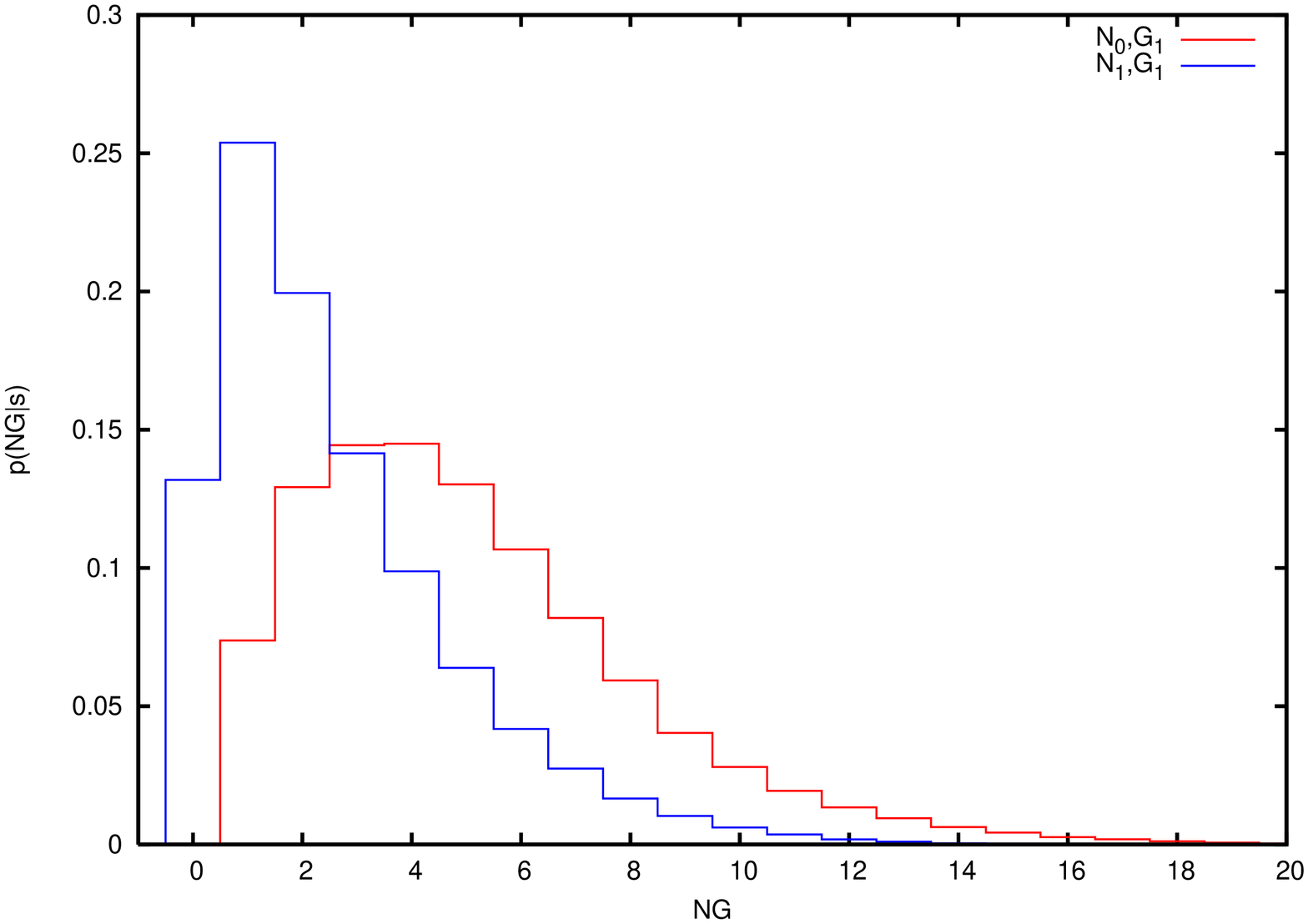}}
  \caption{Plots (a)-(c) show the log evidence for different noise models applied to signal realizations ($s_1$, $s_2$, $s_3$) as a function of the TFV used by $N_0$ and $N_1$.  Models with discernibly higher evidence are preferred.  Plot (d) shows the distribution of the ``hot-pixel'' number used by $G_1$ for the signal realization $s_3$.  Our interpretation of these results is discussed in the text.}
  \label{noise model evidence}
\end{figure*}

Each data set was analyzed using the four different combinations of noise models described above:  $[N_0, G_0]$, $[N_1,G_0]$, $[N_0,G_1]$, and $[N_1,G_1]$, for noise blocks of different time frequency volume (TFV).
Histograms showing the evidence for each model as a function of $\log_2(\rm{TFV})$, as well as the posterior
for the hot-pixel number $n_G$ for $s_3$, are displayed in Figure~\ref{noise model evidence}.

We will first address the dependence of the model evidence on TFV.  For $s_1$ and $s_3$ the data were dominated by Gaussian noise, while $s_2$ contained a significantly ``loud'', short duration, glitch.  Intuitively, noise models with fewer degrees of freedom are preferable when the noise level from block to block hardly strays from the theoretical prediction ($s_1$, and to some extent, $s_3$), while models with more flexibility should be favored when a large deviation from Gaussian noise exists.  This presumption is supported by the results as we see the evidence improves with increasing TFV (decreasing number of noise parameters) for $s_1$ and $s_3$, and diminishes (considering models without glitch fitting ($G_0$)) for $s_2$. 

This study has also allowed us to make some conclusions about selecting between these different models for non-Gaussian noise.  We anticipated that the favored model would most qualitatively resemble the noise simulation, to wit, [$N_0$, $G_0$] for $s_1$, [$N_0$, $G_1$] for $s_2$, and [$N_1$, $G_0$] for $s_3$.  This prediction, however, is not entirely supported by the results.  

For $s_1$ we see that [$N_0$, $G_0$] is on either equal footing (for small TFV), or mildly disfavored (at large TFV), when competing with [$N_1$, $G_0$] and [$N_0$, $G_1$].  While [$N_0$, $G_0$] is the noise model which most faithfully represents how the noise for $s_1$ was generated, it's inflexibility when compared to [$N_0$, $G_1$] and [$N_1$, $G_0$] makes it less well suited to cope with a particular noise realization.  It should be pointed out, however, that the differences in evidence in this example are not significant in a model-selection sense.

For $s_2$ we see a strong preference for models which include glitch fitting ($G_1$).  This is to be expected when the data contain a large amplitude noise impulse, while the $G_0$ models are more adept at describing instrument noise with more frequent, lower amplitude, and uncorrelated noise excursions.

The data simulation which is arguably the most relevant to our ambition of including this work in the analysis of current gravitational wave data would be $s_3$.  Here we have a large number of glitches (300) injected into the data with SNRs drawn from a distribution calibrated to match the frequency of black hole ring-down triggers in S5 data.  Our motivation for this realization is to produce a signal that is most like the actual interferometer data.  The resulting noise ``looks'' predominantly Gaussian, as the vast majority of the glitches are well below the normal instrument noise level.  However, when the data are analyzed with the noise models we see a generic disfavoring of the Gaussian model [$N_0$, $G_0$].  Furthermore, we are pleasantly surprised to see that [$N_0$, $G_1$] performs as well as [$N_1$, $G_0$] in this example even though  [$N_1$, $G_0$]  is qualitatively best  suited for this noise simulation. 

The final point of interest regarding the performance of the different noise models is the generally underwhelming performance of [$N_1$, $G_1$].  The expectation going in to the study was that it would outperform its counterparts on $s_3$, as it was best suited to handle both the rare impulsive events, and the overall non-Gaussian component from the superposition of many small-amplitude glitches.  To explain why it generally underperforms when compared to the other two non-Gaussian models we refer to Fig.~\ref{noise model evidence}d.  The plot shows the posterior distribution functions for the number of hot pixels $n_G$ used by the glitch fitting.  The Gaussian likelihood model, [$N_0$, $G_1$] (red), has a peak in the posterior at $n_G=4$, while [$N_1$, $G_1$]'s peak (blue) is closer to zero, at $n_G=1$.  If the noise model uses a Gaussian distribution for the likelihood some number of hot pixels are required to achieve a sufficiently Gaussian residual, and $G_1$ has the flexibility to do so.  However, if the likelihood accounts for these non-Gaussian excursions by having more weight at large $\sigma$, fewer hot pixels are needed.  $N_1$ is, by itself, an accurate representation of the noise and adding $G_1$ to the modeling introduces additional degrees of freedom without a substantial benefit to the overall fit, making it a less attractive choice in a model selection sense.  

Another unanticipated result can be seen in Fig.~\ref{noise model evidence}d.  We generically see that the posterior distribution on $n_G$ is always peaked away from zero.  This can be understood by studying the Hasting's ratio for transitions from $n_G=0$ to $n_G=1$.  Under such transitions, 
\begin{equation}
H_{0 \rightarrow 1} = \frac{p(s|1)}{p(s|0)} \frac{p(i,j|1)}{p(i,j|0)} \frac{p(\gamma_{ij}|1)}{p(\gamma_{ij}|0)} \frac{q(i,j|0)}{q(i,j|1)} \frac{q(\gamma_{ij}|0)}{q(\gamma_{ij}|1)} 
\end{equation}
with
\begin{eqnarray*}
q(i,j|0) = 1&:& \text{there is only one pixel to remove in} \nonumber \\ &&\text{a transition from $1\rightarrow0$.}\nonumber \\
q(i,j|1) =1/N &:& \text{there are $N$ equally likely pixels} \nonumber \\ &&\text{to include in a transition $0\rightarrow1$.} \nonumber \\
p(i,j|0) = 1 &:&\text{there is only one configuration of}\nonumber \\ && \text{pixels if none of them are ``hot.''}\nonumber \\
p(i,j|1) = 1/N &:&\text{there are $N$ configurations for}\nonumber \\ && \text{a single ``hot'' pixel.}\nonumber \\
p(\gamma_{ij}|m) = q(\gamma_{ij}|m) &:&\text{by construction.}\nonumber
\end{eqnarray*}
and $H_{0 \rightarrow 1}$ reduces to the likelihood ratio.  Therefore, as long as the inclusion of a single hot pixel does not \emph{increase} the residual, $p(s|1)>p(s|0)$, $H_{0 \rightarrow 1} > 1$, and the transition to
$n_G=1$ is accepted.

The results of these noise model comparisons lead us to the following conclusion:  While more than one model is preferred under different circumstances (noise/glitch realizations), the model which uses a Gaussian likelihood along with glitch fitting was \emph{never} seen to be \emph{disfavored} in comparison to the other models under consideration.  Given this generic utility, we will adopt [$N_0$, $G_1$] as our default noise model while turning our attention to the model selection challenge in which we are more fundamentally interested:  Distinguishing between a \emph{gravitational wave signal} and non-Gaussian noise.

\section{Modeling the Gravitational Waves}
\label{waveform}
We take as our test signal the inspiral of a binary system composed of two stellar-mass black holes.  The full gravitational waveform from the inspiral of binary black holes is composed of an inspiral, merger, and ringdown phase.  Analytic solutions exist for the inspiral and ringdown waveforms courtesy of post-Newtonian and black hole perturbation theory, respectively.  Recent advances in Numerical Relativity have led to the simulation of black-hole mergers, allowing (for particular combinations of mass and spin) the construction of full waveforms.  Analytic models of the full waveforms, calibrated against the numerical simulations, provide accurate waveforms which cover the entire signal (see {\it e.g}~\cite{Buonanno:2007pf, Pan:2009wj}).  While a true analysis pipeline should utilize all of the available signal, for this proof-of-principle effort we will only simulate the inspiral phase of waveform both for our injections as well as our analysis.

While neglecting the post-inspiral part of the waveforms, we apply additional simplifications by ignoring spin effects, as well as any orbital eccentricity.  What remains is a model of the waveforms which is characterized by nine quantities written as components of the signal-model parameter vector:
\begin{equation}
\vec{\lambda}\rightarrow(m,\mathcal{M},t_c,\log D_L,\sin\delta,\alpha,\cos\theta_L,\phi_L,\varphi_c)
\end{equation}
where $m$ and $\mathcal{M}$ are the total and chirp mass of the binary system, $t_c$ is the time when the binary coalesces (more accurately, when the post-Newtonian approximation to the frequency diverges), $D_L$ is the luminosity distance to the system, $\delta$ and $\alpha$ are the declination and right ascension as defined on the celestial sphere, $\theta_L$ and $\phi_L$ are the polar coordinates of the angular momentum vector for the binary, and $\varphi_c$ is the GW phase at coalescence. We will further simplify the waveform by only considering the quadrupole radiation, ignoring higher harmonics.  For parameter estimation studies which incorporate the details that we have ignored see, for example,~\cite{Lang:2007ge, Porter:2008kn, McWilliams:2009bg, Key:2010tc}.

\section{The Detection algorithm} 
\label{detection}
Armed with models for the noise and the signals, we now wish to test their relative distinguishability on simulated data using the PTMCMC detection algorithm.  We will consider four models: 
\begin{eqnarray}
N_0, G_0, B_0:  s(t,f) &=& n(t,f), \nonumber \\
N_0, G_0, B_1:  s(t,f) &=& n(t,f) + h(t,f), \nonumber \\
N_0, G_1, B_0:  s(t,f) &=& n(t,f) + g(t,f), \nonumber \\
N_0, G_1, B_1:  s(t,f) &=& n(t,f) + g(t,f) + h(t,f), 
\end{eqnarray}
and for each, thoroughly resolve the PDF of the model parameters and calculate the evidence.  We will consider the prior odds between the models to be unity so that differences in model evidence are equal to differences in the posterior probability for each model.  Henceforth, we will omit writing $N_0$ as part of the model identification as it is common in all scenario's considered here.

The detection algorithm is broadly divided into three phases.  For each model under consideration the algorithm performs a:
\begin{itemize}
\item \emph{Search}:  To locate the regions of high probability in parameter space.
\item \emph{Characterization}:  To globally sample the posterior distribution for the model parameters. 
\item \emph{Evaluation}:  To calculate model evidence and determine which (if any) of the models are favored.
\end{itemize}
The stages of the algorithm are linked automatically, making this an ``end-to-end'' analysis pipeline for matched-filtering searches.  Generally, the search phase sacrifices detailed balance in favor of rapid convergence, and uses a modest number of parallel chains ($\sim10$).  Information from the search is used to construct informed proposal distributions for the characterization, where we must take care that the chains are Markovian.  During the characterization phase we employ a stronger dose of parallel tempering ($\sim 30$ chains) to thoroughly sample the evidence integrand.  When using thermodynamic integration of the evidence, the evaluation
phase occurs in post-processing as all of the necessary information is acquired during the
characterization phase.

The above description of the detection algorithm is a very general outline of the procedure.  What follows is a more detailed look into the implementation of each step for this study.

\subsection{The Glitch Search}
We first analyze the data from each detector individually with the glitch-only model ($G_1$, $B_0$), to establish the range of hot pixels over which the model posterior has significant support.  The noise in each detector is independent, as is the noise model (including glitch fitting).  Therefore, we need not perform this phase of the analysis simultaneously across the network.  

While the $[G_1$, $B_0]$ chains are running in their post-burn-in stage, the state of the glitch model (i.e., the number, location, and amplitude of non-zero wavelet coefficients used in the glitch fitting) is periodically stored so that we retain a population of samples from the glitch search chains.

This ``glitch cleaning'' approach would not work if the data contained un-modeled bursts of gravitational waves as the wavelets efficiently fit the astrophysical signal (as is demonstrated in the wavelet-domain LSC burst search algorithms).  We can perform the analysis in stages here because the power in a single bin of data from a black hole binary signal is swamped by the instrument noise, and it is only with matched filtering that we can effectively elevate these signals above the noise.  The glitch model is therefore ``blind'' to the BH waveform in the data, and is left to clean the noise down to some Gaussian residual. In \S~\ref{discussion} we describe an
improved approach that combines the glitch fitting in each detector with a search for coherent power
across the detector network.

\subsection{The Gravitational Wave Search}
The goal of the GW search phase is to locate the modes of the posterior distribution function.  Straight forward implementations of MCMCs converge slowly -- impractically so when exploring large spaces of high dimensionality.  Several methods exist for optimizing a search using the Metropolis-Hastings algorithm as the driver for the exploration~\cite{Cornish:2005qw, Crowder:2006eu, Cornish:2007jv, Brown:2007se, Gair:2008, Robinson:2008fb, Trias:2008dc,Key:2008}.  The most effective are those which either violate detailed balance and/or sample from a biased target distribution (e.g. simulated annealing, F-statistic extremization, ``mode-hopping'' proposals, etc.).  There do exist enhancements which preserve the independence of the chain samples, such as delayed rejection, but we have found the implementation of such methods unnecessarily complicated.  

Our approach is to use ``illegal'' search techniques to locate the regions of high posterior weight and to use the biased PDF from this search as one of the proposal distributions used to sample from the target distribution.  Specifically, we maximize the likelihood over a subset of the extrinsic parameters, $\{t_c, \varphi_c, D_L\}$ using the correlation of the template with the data.  This maximization must be done in the Fourier domain and assumes gaussian noise.  The search is therefore performed in frequency space and in the absence of noise fitting.  Consequently, the search-phase inner products use the initial LIGO and Virgo theoretical noise PSD.  

The disadvantage to this Fourier-domain, biased search phase lies in our inability to do any glitch fitting during this stage.  To prevent the searching chains from spending time fitting to glitches, we make use of the stored states from the glitch removal phase.  A state of the (already finished) glitch chain is randomly chosen, and subtracted from the data.  The subtraction is done in the time-domain, and occurs periodically, about once every 1000 iterations, so the search phase sees the residual from different realizations of the glitch-only analysis.  Performing this maneuver while maintaining detailed balance would be a challenge but because we are merely localizing the modes of the posterior (and not claiming that we are accurately sampling the PDF), we can get away with these violations.

\subsection{Characterization and Evaluation}
To characterize the signal and glitch model, and evaluate each model's evidence, we repeat the analysis with the full implementation of parallel tempering and noise fitting while satisfying detailed balance and sampling from the target distribution.  We \emph{greatly} expedite this phase of the analysis by including the proposal distribution that was constructed from the (biased) samples of the GW search phase.  Recall that the proposal distribution, by construction of the Hastings ratio, can not bias the chain's sampling of the target distribution.  Even though the search-chain samples were obtained ``illegally', they \emph{can} be used to help move the chains between modes of the target distribution.

To form this efficient proposal distribution, we sort the GW search chain parameters into a joint histogram with bin widths (in each parameter direction) equal to some multiple of the standard deviation for that parameter as estimated by the Fisher Information Matrix.  The Fisher Information Matrix is evaluated at the search chain's Maximum a Posteriori (MAP) parameters.  The technique of constructing a proposal distribution for a high-dimensional space which has been sparsely covered by a Markov chain was described with more detail in~\cite{Littenberg:2009bm}.  Binning the search chains into a proposal distribution can most likely be made more efficient by using binary space partitioning algorithms -- a technique that has previously been studied for use in LIGO/Virgo analyses to approximate the posterior distribution from a
Markov chain~\cite{x}.
 
Provided the resulting 9-dimensional histogram built out of the search-chain samples is normalized and allows access to the entire prior volume it is a perfectly valid proposal distribution from which we can draw new parameter combinations (taking care to accurately calculate the $q(x|y)/q(x|y)$ term in the Hastings ratio).  This proposal distribution, although biased by the likelihood maximization from the search phase, is a sufficiently accurate approximation to the true PDF to improve the mixing of the chains during the characterization phase.  The use of parallel tempering, coupled with the pilot exploration of the posterior, leads to rapid (less than $10^3$ iterations) convergence and supplies further assurance that the chain is globally sampling the target distribution.  

We further expedite matters by restricting the number of hot pixels in the glitch model to those which had significant weight in the search-phase posterior.  Effectively, we are reducing the number of glitch models under consideration and neglecting those which would contribute negligible weight to the evidence.

Upon the completion of the characterization chains, the average likelihood for each chain is computed, and the evidence is calculated using a simple trapezoid integration over the inverse temperature.

\subsection{Prior distributions}
We use uninformative (flat) priors on all of the signal model parameters with angular variables taking the full allowed range.  We allow for the individual constituents of the binary to have masses ranging from 1 M$_{\odot}$ to 30 M$_{\odot}$.  The distance to the binary is constrained between $10^{-1}$ and $10^3$ Mpc.  Because the masses are scale parameters we use uniform distributions in the $\log$ of the chirp and total mass.  Each time-frequency block for which a noise parameter is assigned contains $N_{TFV}=1024$ pixels.  

For more detail regarding the implementation of the MCMC and parallel tempering, refer to the Appendix.

\section{Results} 
\label{results}
The data are simulated by first creating a noise realization as described in \S\ref{noise}, and then injecting a black hole waveform with the luminosity distance tuned to achieve the desired network signal-to-noise ratio ${\rm SNR}=\sqrt{(h\vert h)}$. The injected SNR is computed using the theoretical, Gaussian
noise level, so the actual SNRs will be somewhat lower since additional noise is injected by way of
the glitches.
We will study the performance of the detection algorithm on two noise simulations qualitatively similar to $s2$ and $s3$ from \S\ref{noise} over a three-detector network which includes both four kilometer LIGO interferometers plus Virgo. 

\begin{figure}
\includegraphics[angle=0, width=0.5\textwidth]{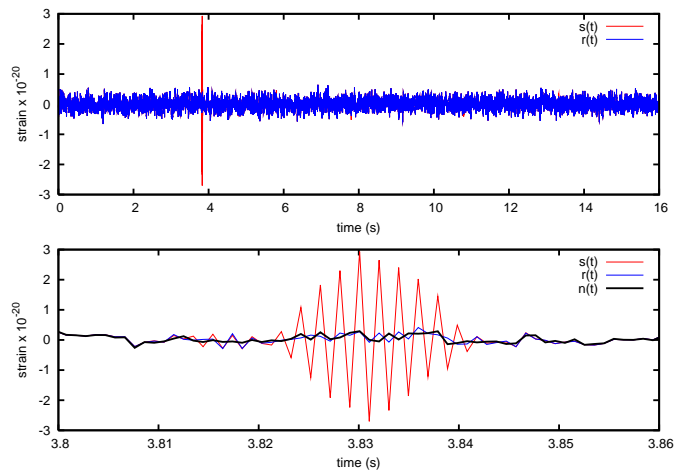} % requires the graphicx package
\caption{A typical residual from the glitch search phase.  The top panel shows the full 16 seconds of
Virgo data while the bottom is a zoomed in view of the glitch.  The red line is for the raw data
$s(t)$, while the blue line is the residual $r(t)$ from a randomly selected state of the
glitch-search chain.  Included in the lower panel is the injected Gaussian noise $n(t)$
demonstrating the accuracy of the glitch removal.}
\label{Results:residual}
\end{figure}

\subsection{Example 1:  Network of Detectors, Single Loud Glitch}\label{Results:ex1}
For this example, the BH injections had SNRs between 6 and 12.  The glitch injection contained a single glitch atom, injected into the simulated Virgo data, with a SNR of 100.  The time-frequency location of the glitch was chosen to overlap with the BH waveform.  

\begin{figure}[htbp]
   \includegraphics[angle=0, width=0.5\textwidth]{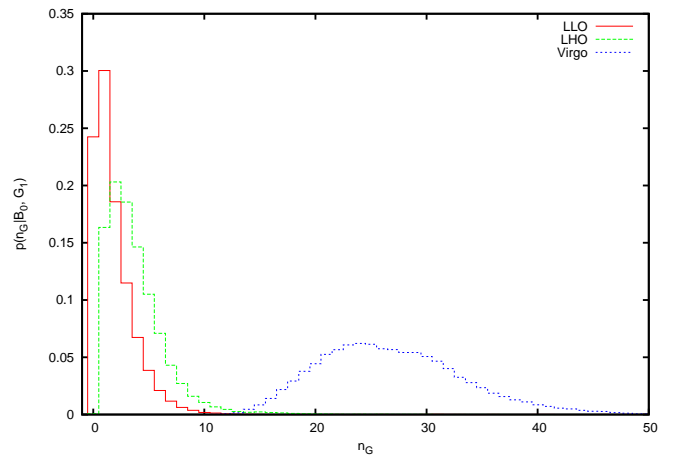} % requires the graphicx package
   \caption{ PDF of $n_G$ for stationary, Gaussian noise with a single ``loud'' glitch injected into the Virgo data.  The RJMCMC automatically prefers small values of $n_G$ for the Gaussian noise, and $\sim$ 25 glitch wavelets to fit the large amplitude noise excursion seen in Figure~\ref{Results:residual}.}
   \label{Results:glitchpdf1}
\end{figure}

\begin{figure}[htbp]
   \includegraphics[angle=270, width=0.5\textwidth]{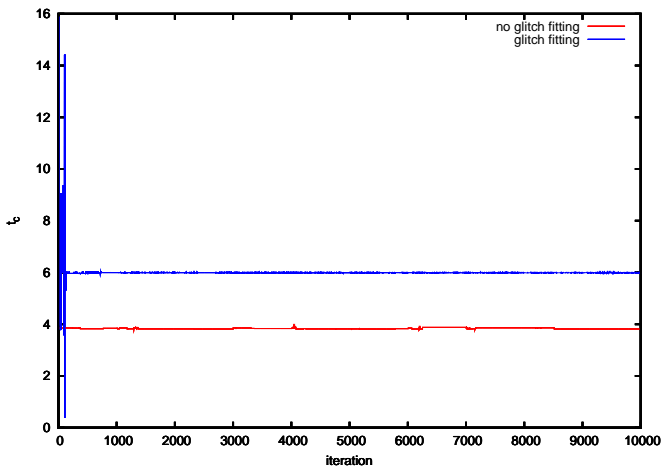} % requires the graphicx package
   \caption{Time-to-coalescence search chains with (blue) and without (red) glitch fitting for Example 1.
Without glitch fitting the search locks onto a glitch in the Virgo detector (injection time $\sim 4$ s)
while with glitch fitting the search locks onto the signal injection ($t_c\sim$ 6 s).  The GW signal
was injected with SNR of 12.  The prescription for the glitch fitting is described in \S\ref{detection}.}
   \label{Results:search1}
\end{figure}

Figure~\ref{Results:residual} shows the accuracy with which the glitch is removed from the data.  The top panel displays the full 16 seconds of Virgo's time-domain data (red) and the residual (blue) for a typical state of the chain.  The lower panel zooms in to the region around the glitch, and includes the injected Gaussian noise (black).  The RJMCMC glitch fitting procedure was able to parsimoniously remove the non-Gaussian component of the noise without significantly affecting the rest of the data. 

\begin{figure*}[htbp]
   \includegraphics[angle=0, width=\textwidth]{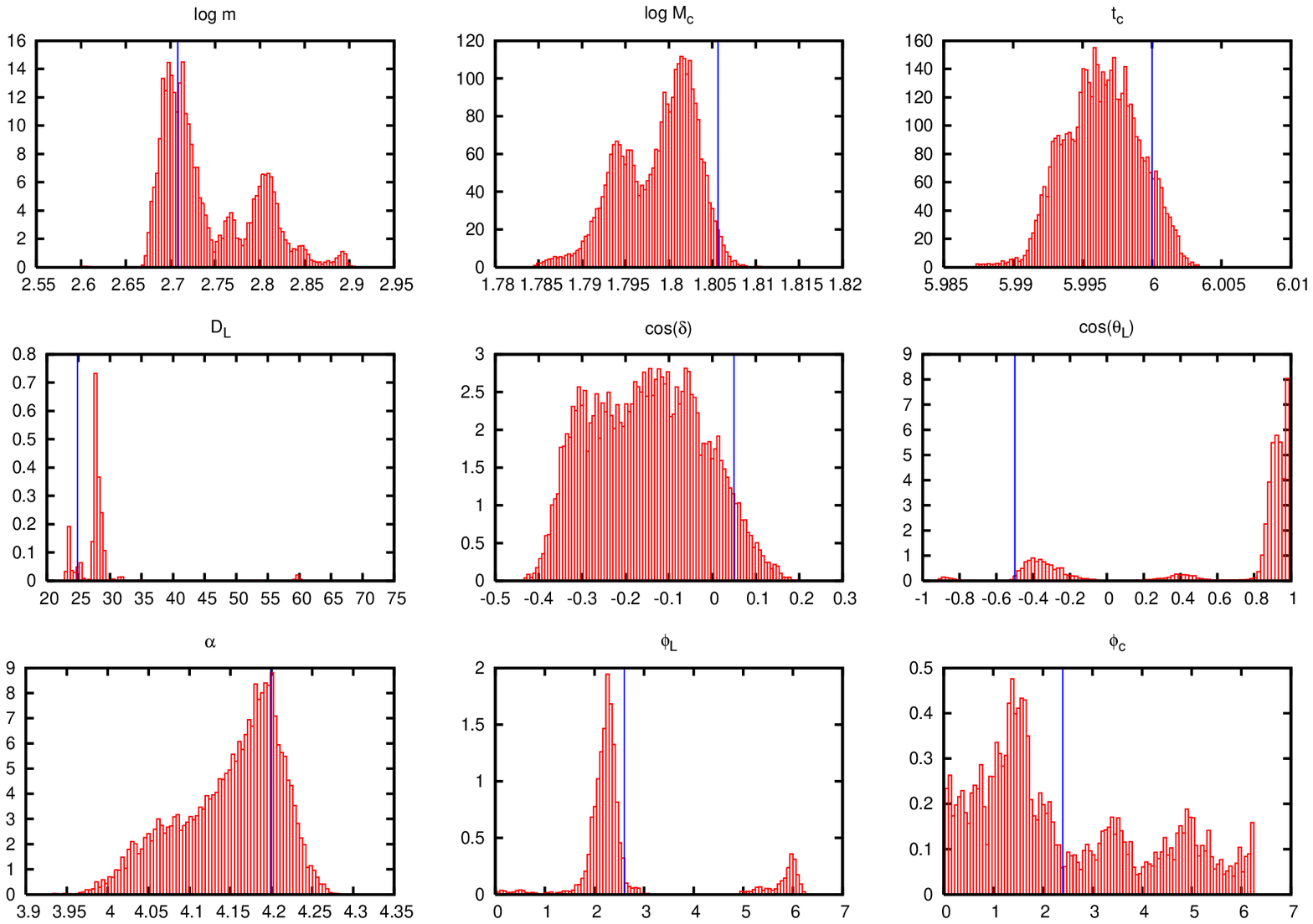} % requires the graphicx package
   \caption{Marginalized PDFs of $\vec{\lambda}$ for SNR~$=12$ from Example 1.  The vertical line marks the injected waveform's values.  Notice the multi-modal structure of the posteriors, illustrating the challenge these signals pose for any analysis method.}
   \label{Results:pdf1}
\end{figure*}

The distributions of the number of hot pixels ($n_G$) used by the RJMCMC glitch model are shown in Figure~\ref{Results:glitchpdf1}.  The value of $n_G$ at the peak of these distributions is the glitch model with the highest evidence.  The favored models for the LIGO Livingston and Hanford detectors (LHO and LLO) are for low $n_G$ as the simulated noise from those instruments was stationary and Gaussian, while we see clear evidence for a non-Gaussian feature in the Virgo data.  For the characterization/evaluation studies we restrict $n_G$ in each interferometer to those in Figure~\ref{Results:glitchpdf1} which have non-zero probability.  For discussion of why the LLO and LHO histograms are not peaked at zero see \S\ref{noise}.

The black hole search was run on data with the glitch already removed as described in \S\ref{detection}.  For demonstrative purposes, we also ran a search on the ``raw'' data still containing the glitches.  Figure~\ref{Results:search1} shows the time-to-coalescence chains for the searches performed on the data with the BH injection at SNR of 12.  The search templates on the raw data had $t_c$ values consistent with the time of the glitch injection (completely missing the BH signal), while the chains exploring the residual data found the correct value of the merger time.  For the other data sets,  the injected source parameters were successfully extracted from the residual data for injections above signal-to-noise ratios of $\sim 8$.

Upon completing the search chains, we bin the samples from the BH search into a 9D histogram to use as a proposal distribution (as described in \S\ref{detection} and Ref.~\cite{Littenberg:2009bm}) and reanalyze the data using the four models outlined in the previous section.  During this characterization phase the noise and signal
parameters are updated together, and care is taken to satisfy the detailed balance condition. 

Figure~\ref{Results:pdf1} shows the posterior distribution functions for each of the BH signal parameters, marginalized over all other parameters and noise models, for the SNR~$=12$ signal injection.  Vertical lines denote
the injected parameter values, all of which lie in the supported region of the posterior.  The familiar multi-modal structure of these posterior distributions is in evidence.

\begin{figure}[htbp]
   \includegraphics[ width=0.5\textwidth]{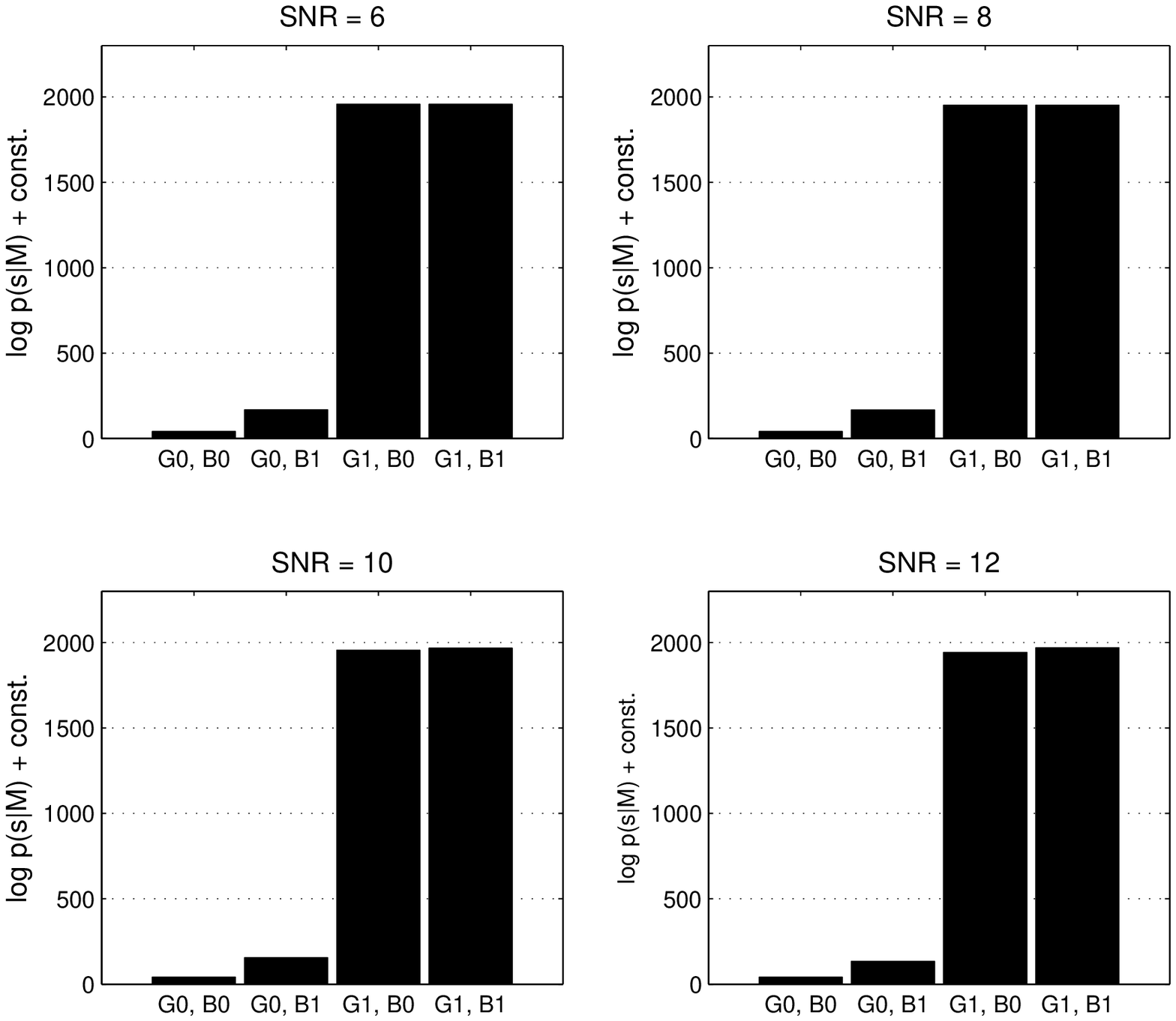} % requires the graphicx package
   \caption{Log evidence for each model under consideration for Example 1.  The evidence is highest for
the glitch-fitting models. The BH plus glitch model $[G_1, B_1]$ begins to distinguish itself from
the glitch model $[G_1, B_0]$ above injected SNR of 8.  Had we naively assumed a Gaussian noise model,
the evidence calculation would have erroneously indicated a detection for arbitralily weak signal
injections (false positives in the usual parlance).}
   \label{Results:evidence1}
\end{figure}

The evidence for each model is calculated using thermodynamic integration and the results of these calculations are shown in Figure~\ref{Results:evidence1}.  The evidence calculation is un-normalized, so the value for each model is of no consequence.  What matters is the relative evidence between the different descriptions of the data.  For all cases, the models with glitch fitting ($G_1$) dominate over those without ($G_0$).  For the $G_1$ models, we begin to see significant support for the black-hole model ($B_1$) at SNR of 10.   There is never significant support for the noise-only model because we set the prior range on $D_L$ to extend well beyond the range of the detectors, i.e., included in our GW model are signals which are undetectable.  The model selection scheme should not distinguish between noise only or noise plus a gravitational wave of (near) zero amplitude.  Had we adopted some minimum SNR cutoff on the luminosity distance, we would have seen the noise model being favored at low SNR.  A similar study was performed for galactic binaries in Ref.~\cite{Littenberg:2009bm}.

\begin{figure}[htbp]
   \includegraphics[ width=0.5\textwidth]{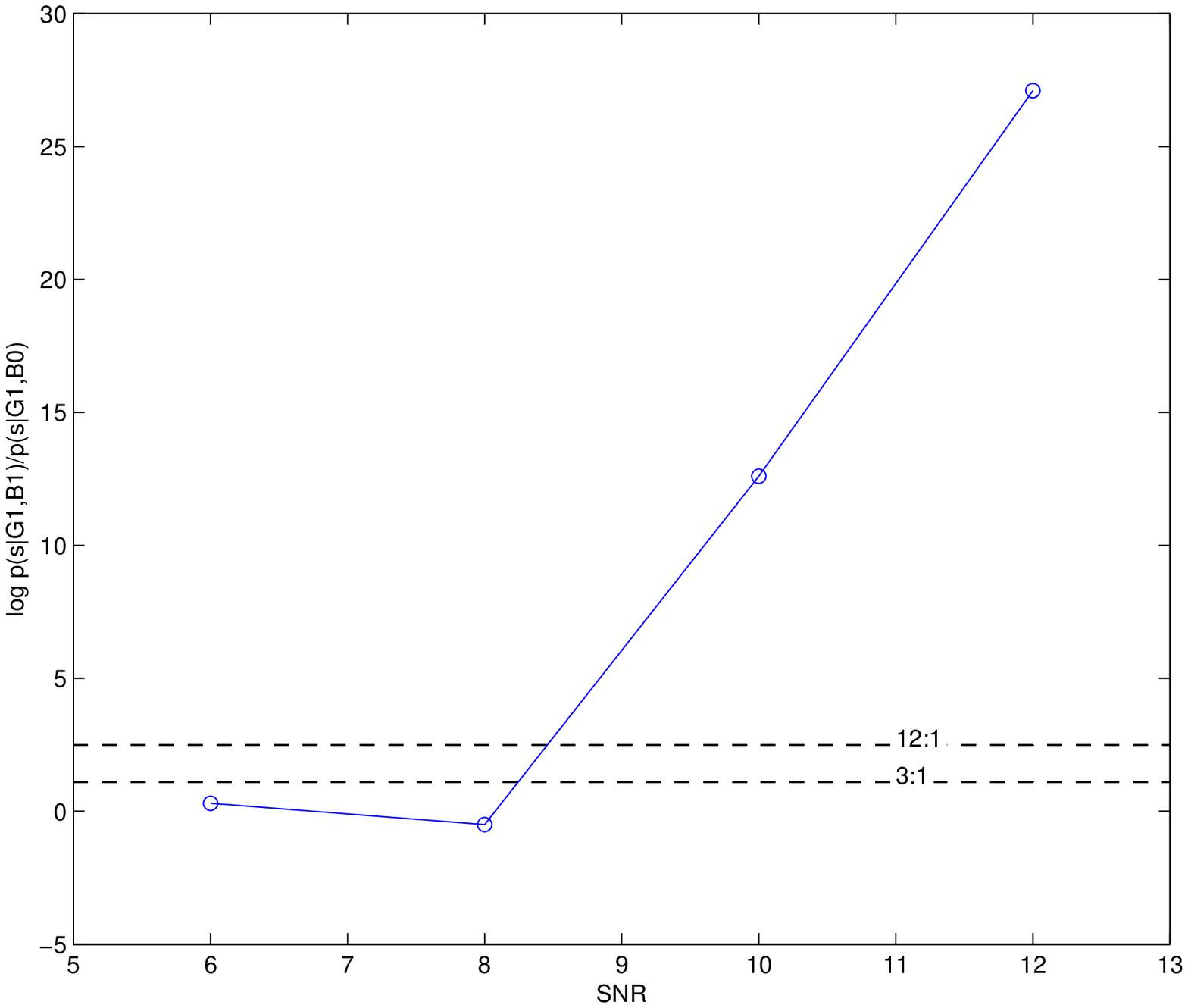} % requires the graphicx package
   \caption{Log Bayes factor for models $[G_1, B_1]$ vs. $[G_1, B_0]$ for Example 1.  The GW-signal
model is favored when the injected signal has SNR above $\sim$8.}
   \label{Results:bayes1}
\end{figure}

The question of ``detectability'' is more clearly demonstrated by looking at the Bayes Factor
\begin{equation}
{\cal B}_{1,0} = \frac{p(s|G_1,B_1)}{p(s|G_1,B_0)}
\end{equation}
shown in Figure~\ref{Results:bayes1}.  Because of our choice to use uniform priors on the different models, ${\cal B}_{1,0}$ is equivalent to the odds ratio for favoring the BH model to the noise-only model.  The dashed lines represent odds ratios of 3:1 and 12:1 which are historically taken as different ``confidence'' intervals in Bayesian model selection.  For Bayes factors between 3:1 and 12:1 model 1 is weakly supported over model 0, while above 12:1 the support for model 1 is considered strong.   We find that by SNR $=10$ there is a clear preference for the signal model which then increases exponentially with the gravitational wave amplitude.  

Perhaps more importantly, we do not see false positive detections (i.e., the BH model fitting to the glitch with significant evidence) when the BH amplitude is too low to be detected.  This contrasts with the models that treat the noise as Gaussian ($G_0$) where we see strong support for the BH model even though the GW chains never located the injected signal (see Figure~\ref{Results:evidence1}).  

\subsection{Example 2:  Network of Detectors, Glitch Distribution}

\begin{figure}
\includegraphics[angle=0, width=0.5\textwidth]{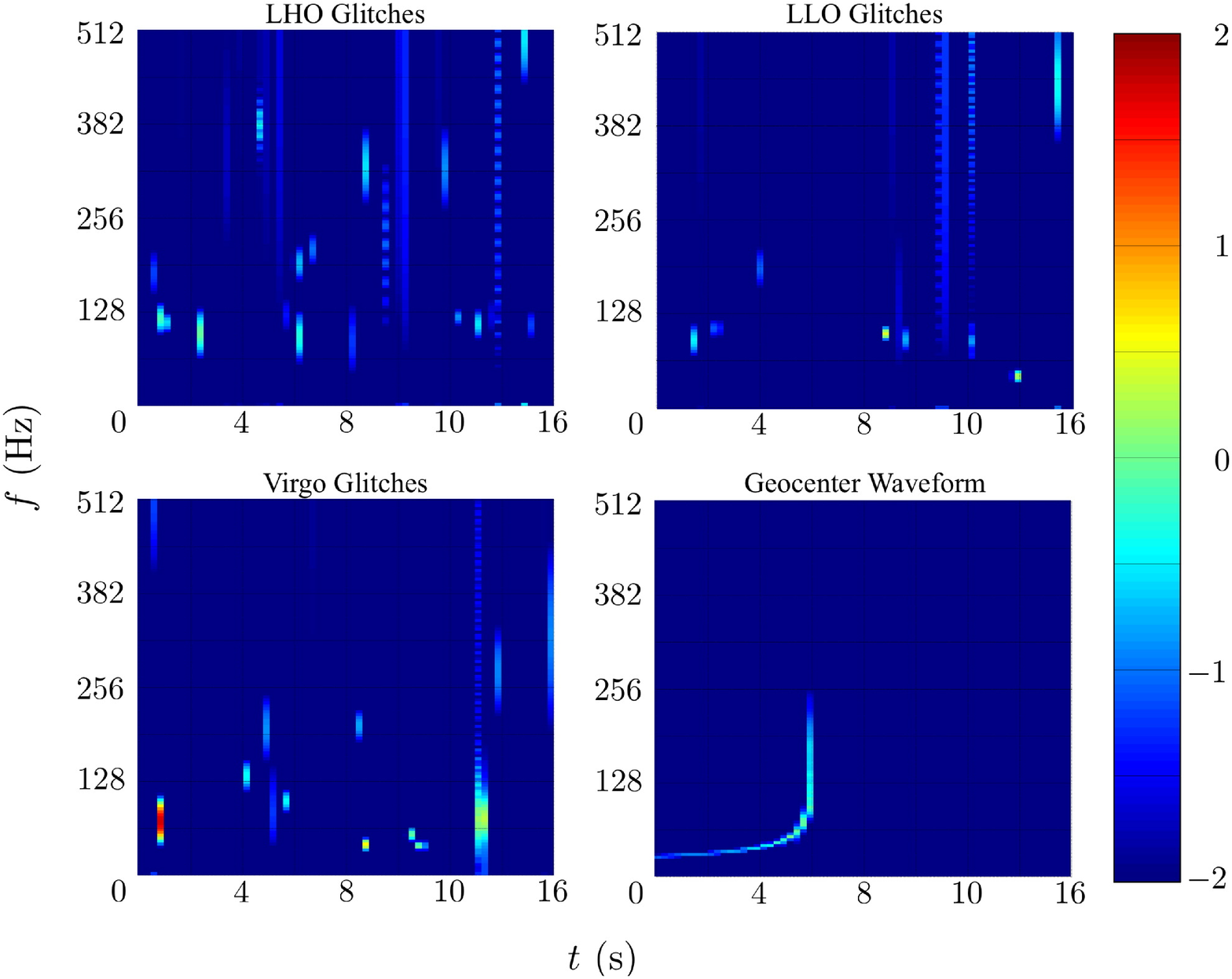} % requires the graphicx package
\caption{Spectrograms showing the glitch injections into the three detectors and the geocenter
gravitational waveform. The spectrograms have been whitened using the theoretical LIGO/Virgo noise spectra.   The color map is on a logarithmic scale, and represents the Fourier power.}  
\label{Results:glitches}
\end{figure}

\begin{figure}
\includegraphics[angle=0, width=0.5\textwidth]{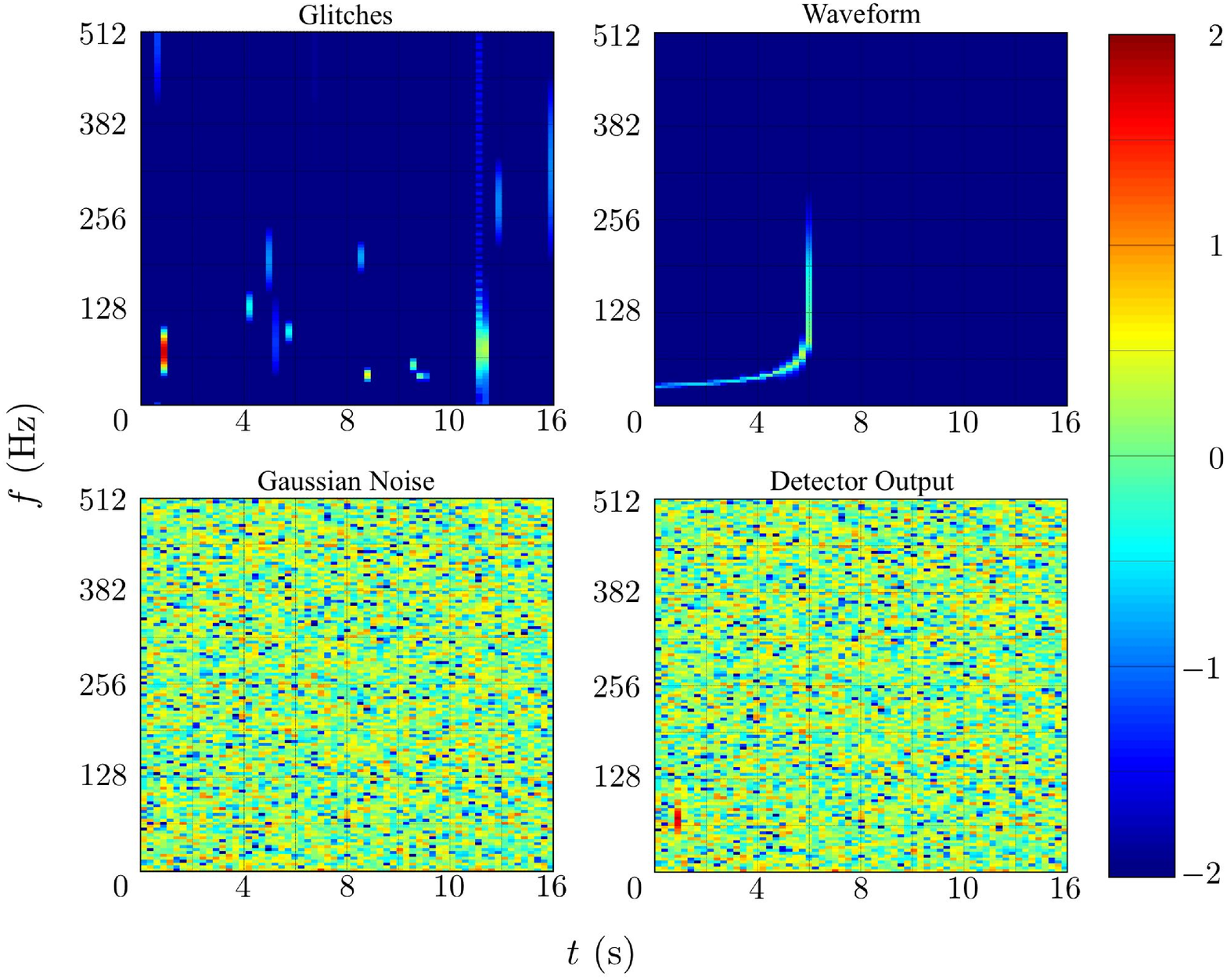} % requires the graphicx package
\caption{Spectrograms showing the various components that make up the simulated data in the Virgo
detector for Example 2. The spectrograms have been whitened using the theoretical Virgo noise spectrum.
The loud, low frequency glitch at $t\sim 1$ second caused havoc for noise models that did not
include glitch fitting.  The color map is on a logarithmic scale, and represents the Fourier power.}
\label{Results:Virgo}
\end{figure}

Example 1 is a simplified simulation of non-Gaussian noise for the ground based GW detectors.  To further test the effectiveness of our noise-modeling algorithm we simulated new data and re-ran the analysis described in Example 1 but with 100 glitches injected from a distribution calibrated off of the S5 ring-down trigger rate as described in~\S\ref{noise}. Spectrograms of the glitch injections into each detector are shown in Figure~\ref{Results:glitches} and are compared to the waveform injection which is shown using the
same color scale. Figure~\ref{Results:Virgo} shows the various contributions to the detector output for
the Virgo detector, which happened to have the loudest glitch in this noise realization
(at $t\sim 1$ second).

\begin{figure}[htbp]
   \includegraphics[angle=270, width=0.5\textwidth]{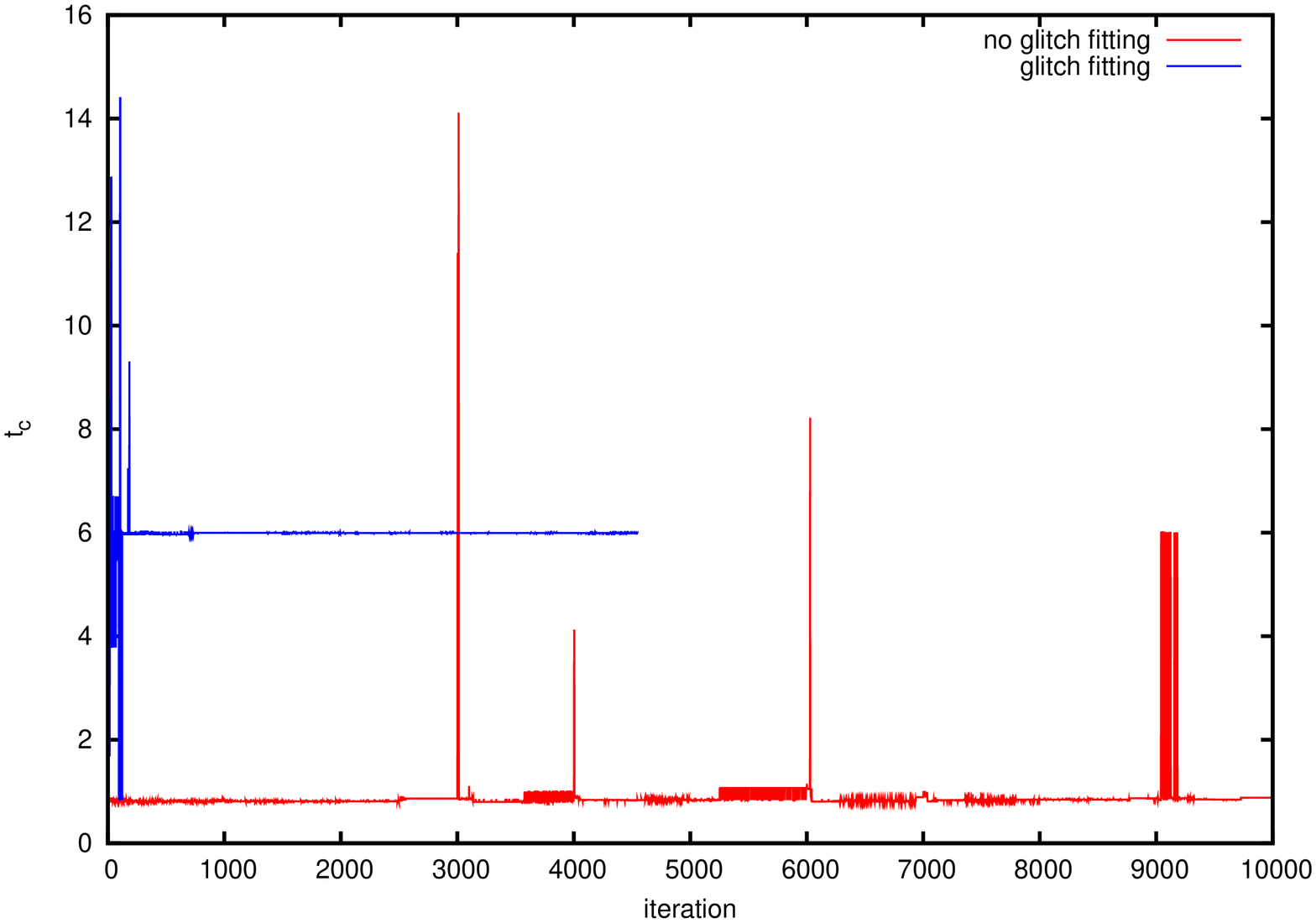} % requires the graphicx package
   \caption{Time-to-coalescence search chains with and without glitch fitting for Example 2.
Without glitch fitting the search locks onto a glitch in the Virgo detector (injection time $\sim 1$ s)
while with glitch fitting the search locks onto the signal injection ($t_c\sim$ 6 s).  The GW signal
was injected with SNR of 12.  The prescription for the glitch fitting is described in \S\ref{detection}.}
   \label{Results:search2}
\end{figure}

\begin{figure}[htbp]
   \includegraphics[ width=0.5\textwidth]{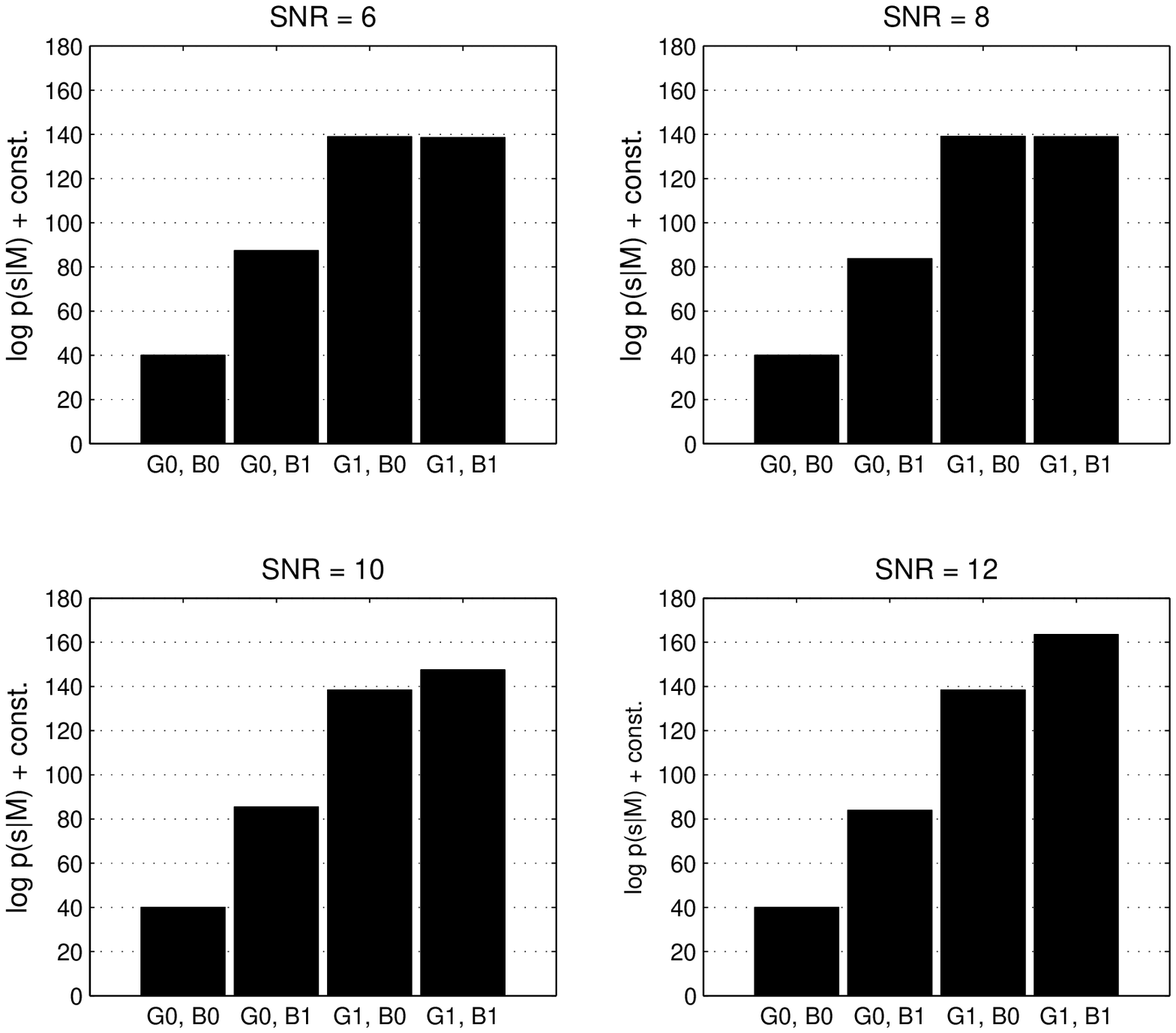} % requires the graphicx package
   \caption{Log evidence for each model under consideration for Example 2.  The evidence is highest
for the glitch-fitting models. The BH model plus glitch model $[G_1, B_1]$ is favored for injected
SNR's above 8.}
   \label{Results:evidence2}
\end{figure}

\begin{figure}[htbp]
   \includegraphics[ width=0.5\textwidth]{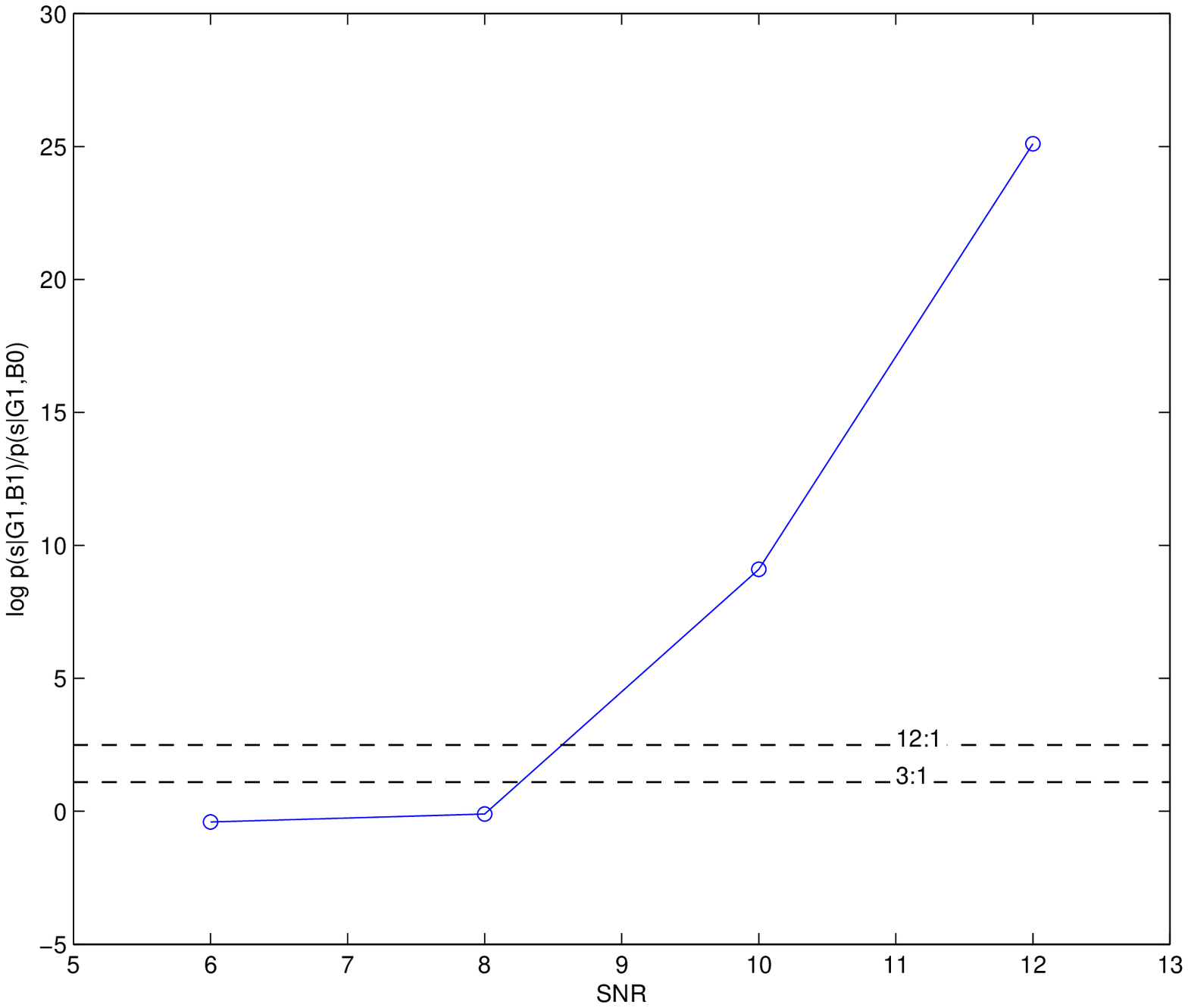} % requires the graphicx package
   \caption{Log Bayes factor for models with glitch fitting, showing the odds that a signal is
present, $[G_1, B_1]$, relative to no signal being present, $[G_1, B_0]$. The signal model
is favored for injected SNR's above 8.}
   \label{Results:bayes2}
\end{figure}

The results for this second example are qualitatively similar to those for the first example, thus the
discussion in this section is kept brief.  Figures~\ref{Results:search2}, \ref{Results:evidence2}, \ref{Results:bayes2} for Example 2 correspond to Figures~\ref{Results:search1}, \ref{Results:evidence1}, \ref{Results:bayes1}
for Example 1.  

Despite the increased complexity of this data simulation, and the increased challenge of extracting the GW signal from the non-Gaussian noise, the results and conclusion made from this example are essentially identical to those from the simpler ``single glitch'' case of Example 1.  The only significant differences between this and the previous test are the degree to which glitch fitting improves the evidence, and to the overall confidence levels of detection.  

The change in evidence between $G_0$ and $G_1$ models is much larger in Example 1, where the injected glitch was a much more predominant feature in the data.  However, the relative difference between evidence with and without glitch fitting, for both examples shown here, demonstrate overwhelming support for $G_1$ models.  

We attribute the slight differences in the detection confidences to the GW signal injections,
which are scaled to achieve a desired ideal SNR~$=\sqrt{(h|h)}$.  Because the data from Example 2
contains, on average, more power in each wavelet layer (due to the additional, irresolvable, low-SNR glitches)
as compared to Example 1, the effective SNR of the glitch injections is lower than the indicated by
the theoretical SNR that appears along the x-axis.

\begin{figure}[htbp]
   \includegraphics[ width=0.5\textwidth]{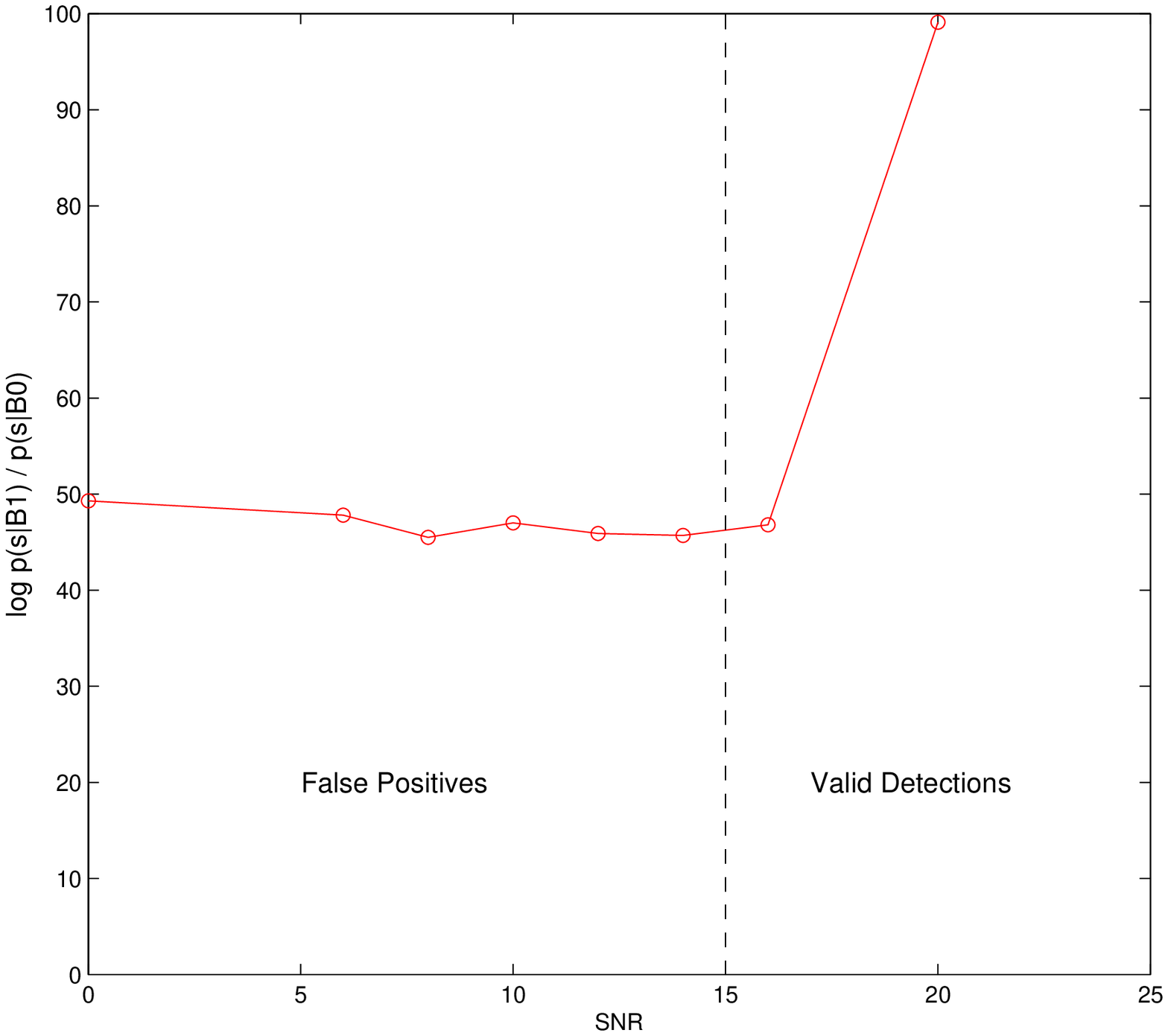} % requires the graphicx package
   \caption{Log Bayes factor for models without glitch fitting, showing the odds that a signal is
present, $[G_0, B_1]$, relative to no signal being present, $[G_0, B_0]$. The signal model is
always favored, even when no signal is injected. The parameters of the recovered signal do
not correspond to those injected until the SNR reaches~$\sim16$.}
   \label{Results:bayes3}
\end{figure}

The evidence ratios for the models that include glitch fitting shown in Figure~\ref{Results:bayes2}
should be compared to the evidence ratios found with no glitch fitting that are shown in
Figure~\ref{Results:bayes3}. Here the dangers of using a Gaussian likelihood function are thrown
into stark relief. A fully convergent Bayesian calculation of the odds ratio favors the signal
model even when no signal is injected! It is not until the injected signal has SNR above 16 that
the recovered parameters actually correspond to what was injected. In contrast, the glitch fitting
model only favors the signal model when the injected signal parameters are recovered. Veitch and
Vecchio~\cite{Veitch:2008ur} have also observed that using a Gaussian noise model to calculate
Bayes factors on glitchy data can lead to ``false positives'', and they have argued that this can be
accounted for by treating the odds ratio as a frequentist statistic, that can be tuned
using signal injections and time slides of the data. While this suggestion is not without merit,
it comes at the cost of having much higher detection thresholds than those demonstrated here when
the noise model includes glitch fitting. It is worth mentioning that the current LIGO/Virgo
inspiral search pipeline~\cite{Abbott:2009tt} would not be fooled into claiming a detection from
the glitches injected in Examples 1 or 2 (events with significant SNR in just one detector do
not make it through to the coincidence test). On the other hand, introducing the capability to
model instrument glitches should help lower the thresholds in these searches.

\section{Discussion} \label{discussion}
The goal of this work was to develop a better likelihood function (i.e., noise model) to account for non-Gaussian and non-stationary features in LIGO/Virgo data, and to apply this new tool to the gravitational wave detection problem.  We studied the performance of three noise models proposed as an alternative to the standard stationary, Gaussian treatment: 
\begin{enumerate}
\item Noise samples are drawn from a Gaussian distribution with a time-varying expectation value. [$N_0, G_0$]
\item The noise distribution is modeled as the sum of two Gaussians such that the bulk of the samples are representative of Gaussian noise but with higher probability for ``large sigma'' events. [$N_1, G_0$]
\item The noise consists of two contributions -- non-stationary, Gaussian samples plus coherent (in time and frequency) ``glitches'' which are modeled by linear combinations of wavelets. [$N_0, G_1$],
\end{enumerate}
and measured their relative success at modeling different non-Gaussian noise simulations by comparing their marginalized likelihoods. 

The most successful of these, [$N_0, G_1$], used a a trans-dimensional Markov chain Monte Carlo algorithm to determine the  most parsimonious number of wavelets used to fit the non-Gaussian constituents of the noise.  This noise model performed as well or better than the others tested for each of the different noise simulations studied (see Figure~\ref{noise model evidence}).  None of these simulations used either of the noise models to generate the noise samples.

The more interesting application of the noise modeling is to use it in conjunction with a gravitational wave search.  We included the glitch-fitting (along with other improvements) into our parallel tempered Markov chain Monte Carlo detection algorithm originally described in~\cite{Littenberg:2009bm}.  The study was performed using two simulations of the interferometer data.  One contained Gaussian noise plus a single ``loud'' glitch intentionally overlapping the injected GW.  The other contained Gaussian noise with 100 glitches distributed across the network having SNRs drawn from a distribution calibrated to match that of single detector's triggers in a LIGO S5 black hole ring-down search.  We find no false positive detections when the glitch modeling is included, and are able to locate the gravitational wave signal injected into data with SNR above~$\sim8$ despite its overlapping with injected glitches.  Detections are claimed using Bayesian evidence ratios calculated via thermodynamic integration (see Figs~\ref{Results:evidence1}, \ref{Results:bayes1}, \ref{Results:evidence2}, and \ref{Results:bayes2}).  

When the glitch fitting is omitted in favor of the non-stationary Gaussian noise model ($N_0, G_0$) the black hole search misses the astrophysical signals and instead attempts to fit to the injected glitches (see Figs.~\ref{Results:search1} and \ref{Results:search2}) until the injected SNR exceeds $\sim 15$.  These false positive scenarios ubiquitously returned very large Bayes factors in favor of detection.  By using a more realistic model for the noise we have mitigated the risk of false-positive detections due to non-Gaussian features and successfully extracted GW signals without throwing away the ``glitchy'' data.

We are encouraged by these demonstrated results and are optimistic that this technique can be used as a foundation for modeling the instrument noise in real data.  There is, however, still much work to be done.  We will proceed down two avenues to improve the utility of our glitch-fitting detection algorithm.  

The first step is to study the performance of these models on actual LIGO/Virgo data in an effort to
further improve the glitch modeling.  In particular, the priors we have constructed for the glitch
parameters (including the number, location in time-frequency space, and amplitudes of the glitches)
were uninformed by the ongoing LIGO/Virgo detector characterization studies.  Folding in this
information should improve the effectiveness of the glitch modeling.

There is no guarantee that wavelets are the best functions with which to decompose the glitches
-- they were chosen in part for their convenience, as they form an orthogonal basis with a diagonal
noise correlation matrix for Gaussian noise.  We found that large glitches required $\sim 25$
wavelet basis functions, whereas the analyses performed by Principe and Pinto ~\cite{Principe:2008bz, Principe:2009zz} were more efficient at matching the features in the data.  Adopting a more
parsimonious basis for fitting the glitches, such as wavelet wavepackets~\cite{Coifman:1992},
could allow the algorithm to dig deeper into the instrument noise and further lower the detection
thresholds.

A weakness of our current implementation is the sequential nature of the initial search phase,
which starts with glitch modeling and is then followed by a search for inspiral signals
in the cleaned data (the characterization phase, by contrast, simultaneously updates the noise
and signal models). This sequential approach is safe so long as the gravitational wave signals
are everywhere below the instrument noise level, but it would be desirable to perform the glitch
removal in way that will not remove loud gravitational wave signals. To this end, we are currently
developing a new version of the algorithm that combines the glitch modeling with a search for
un-modeled gravitational wave signals. Wavelets are used to represent the two gravitational
wave polarizations $h_+$ and $h_\times$ at the Geocenter, and these signals are projected onto
the detector network with appropriate time delays and amplitudes corresponding to the proposed
sky location. It is possible to specify if the signals are un-polarized, or have a
particular waveform polarization (circular, elliptical, linear). If a loud gravitational wave
signal is present in the data, it is more parsimonious for the signal to be assigned to the
Geocenter signal wavelets than the individual instrument wavelets (for the un-polarized search
three detectors are required for this to work, while just two detectors are needed for the
polarized search). The new algorithm runs very quickly since the signals do not have to
be wavelet transformed at each iteration (the timeshifts are done directly in the wavelet
domain). While designed to run as a burst search, the algorithm could be used for data
conditioning prior to other searches by subtracting the best fit glitch model from each
detector and leaving behind any power assigned to the signal model.

\section{Acknowledgments}
We are grateful for input from Bruce Allen, Paul Baker, Kipp Cannon,
Jolien Creighton, Joe Romano and Graham Woan. This
work was supported by NSF grant 0855407.

\appendix
\section*{Appendix:  A Recipe for Parallel Tempering and Thermodynamic Integration}\label{ApA}

Parallel tempering is an effective way to keep Markov chains from locking onto single mode of the target posterior distribution function, with the added benefit of a simple and accurate means of calculating the model's evidence (thermodynamic integration).  Parallel tempering involves running $N_C$ chains simultaneously using a modified Hasting's ratio 
\begin{equation}\label{likelihood}
H_{\vec{\theta}_x \rightarrow \vec{\theta}_y}=
	\left(\frac{ p(s|\vec{\theta}_y,\beta)}{ p(s|\vec{\theta}_x,\beta)} \right)^{\beta}
	\frac{ p(\vec{\theta}_y)q(\vec{\theta}_x|\vec{\theta}_y) }
	{ p(\vec{\theta}_x)  q(\vec{\theta}_y|\vec{\theta}_x) }
\end{equation}
where $\beta = 1/T$ is analogous to an inverse ``temperature'' and takes on an increasing value between 0 and 1 for each chain.  The effect of the exponent on the likelihood ratio in Eq.~\ref{likelihood} is to smooth the topography of the target distribution, making differences between modes of a distribution less dramatic, and broadening the basin of attraction for each mode.  A chain with a high temperature (low $\beta$) will freely move between modes and in the limit where $T\rightarrow\infty$ ($\beta\rightarrow 0$) the likelihood ratio goes to 1 and the chain will sample the prior distribution ($p(\vec{\theta})$).

Hotter chains efficiently sample the full prior volume and, in the event that they locate a region of higher posterior weight, are able to communicate that location to the colder chains without violating detailed balance by using the ``chain swapping'' Hasting's ratio:
\begin{equation}\label{chain swap}
H_{i\leftrightarrow j}=\frac
	{p(s|\vec{\theta}_i, \beta_{j})p(s|\vec{\theta}_{j}, \beta_{i})}   
	{p(s|\vec{\theta}_i,\beta_{i})p(s|\vec{\theta}_{j}, \beta_{j})}.
\end{equation}	
for an exchange between chains $i$ and $j$.  Meanwhile, the colder chains are more apt to thoroughly explore the space around whichever mode it is they sit.  It is as if the hot chains wildly try different solutions, and pass those that offer good fits to the data to the cold chains where they can stored in the history of the chain, and refined with subsequent iterations.

Direct exchange of states between two chains is the simplest way to couple the chains, but not the most efficient.  The use of genetic algorithms (GAs) to produce new solutions (the offspring) from parallel chains (the parents) further increases the benefits of parallel tempering.  

The post-burn-in samples from all of the parallel chains can be used to calculate the evidence for the model under consideration by integrating the average log-likelihood for each chain over $\beta$
\begin{eqnarray}\label{thermo}
\log p(s|\mathcal{M}) &=& \int_0^1 \langle \log p(s|\vec{\theta},\mathcal{M},\beta)\rangle d\beta \nonumber \\
&=&\int_{-\infty}^0\beta\langle \log p(s|\vec{\theta},\mathcal{M},\beta)\rangle d\log\beta.
\end{eqnarray}

By repeating the calculation for different models under consideration, one can calculate the Bayes factor (the ratio of the evidence for two models) or, for a larger number of models, the likelihood distribution (in ``model space'').  If the priors on competing models are uniform, the likelihood distribution is the model posterior up to a normalization constant.

\subsection{Heating schemes}
An efficient implementation of parallel tempering requires the temperature spacing between chains to be large enough that each is free to find it's own stationary state, but small enough that the chains are still frequently exchanging parameters with one another.  At low temperature we tend to err on the side of communication, ensuring that the cold chains are efficiently sampling from modes of the posterior.  This could, depending on the model, mandate very close spacing of the chains.  On the other hand, as the temperature increases the likelihood ratio becomes increasingly less important in the Hasting's ratio, and parameter exchanges become frequently accepted.  In this scenario the chains become over-coupled, preventing them from locating a stationary solution.  This, in turn, affects the accuracy of the evidence integration.  For Gaussian likelihoods, geometric spacing of chain temperatures is most effective.

We use two temperature spacings, $\delta T_h$ and $\delta T_c$, for the hotter and colder chains.  The change-over between the $\delta T$s occurrs at some pre-set temperature $T_*$.  We have typically found that $\delta T_c \sim 1.2 - 1.5$ for the $N_*$ (colder) chains yields adequate mixing.  A more sophisticated choice would be to adjust $\delta T_c$ during the burn-in phase until achieving a desired acceptance rate for the PT proposals. 

The temperature spacing for the remaining $N_C-N_*$ (hotter) chains is calculated such that the hottest chain has temperature $T_{\rm{max}}$ with typical values between $10^2$ and $10^4$.  We elect to fix the highest temperature, instead of the spacing for the hotter chains, so that the evidence integrals for different models occur over the same range regardless of which $\delta T_c$ or $T_i$ we choose for each model.  

To determine $T_*$ we utilize the approximation that (for Gaussian posteriors) the effective SNR of the signal seen by a tempered chain chain is $\rm{SNR_{eff}}\sim\rm{SNR}/\sqrt{T}$.  Because a GW with SNR greater than $\sim10$ will not generally require careful model-selection (assuming we have done an adequate job of modeling the noise) we have used $T_*\sim10$.  This corresponds to a maximum effective SNR at $T_*$ for marginally detectable signals of $10/\sqrt{10}\sim3$ which is sufficiently small to guarantee that hotter chains will not ``see'' the GW signal.  With the $\delta T$s determined, the temperature for each chain in the ladder is
\begin{eqnarray}
T^i &=& \delta T_c^i, \ \ \ \ \ \ \ \ \ \ \ \ 0 \leq i < N_* \nonumber \\
T^i &=& T_*\delta T_h^{i-N_*}, \ \  N_* \leq i < N_C.
\end{eqnarray}

If locating and/or characterizing the modes of the posterior, without calculating the evidence, is the goal of the analysis the temperature ladder need not extend beyond $N_*$.  Hotter chains are primarily sampling from the priors and do not noticeably aid in the convergence of the colder chains (although they are critical in the evidence calculation).  In other words, chains that are to be used for thermodynamic integration have to go out to much larger temperature than those used to produce samples the posterior.  The exact number of chains, and the maximum temperature needed for both scenarios, are tuned on a problem-by-problem basis.  Key diagnostics for determining these settings are discussed later.

\subsection{Proposal distributions}
There are no rules mandating how often, or between which chains, parameter exchanges should be proposed.  We have typically proposed exchanges only between adjacent chains, and done so between each chain once for every iteration of the MCMC.  The parameter-swap proposal will first suggest an exchange between chains $N_C \leftrightarrow N_C-1$ (the hottest, and next to hottest chains), then $N_C-1 \leftrightarrow N_C-2$, and so on, until a proposed parameter switch between chains$1\leftrightarrow 0$ (the next to coldest and coldest chains).  For model selection problems, when the maximum temperature is large, the hotter chains (each effectively sampling from the the prior) accept chain-swaps with excessively high frequency ($> 90\%$).  To prevent the hotter chains from over-coupling with their neighbors we propose exchanges between chains $i$ and $j$, where $T_j < T_i$ with probability $\beta_j$.  The coldest chain has $\beta=1$ and always attempts  parameter exchanges during the PTMCMC proposal while hotter chains attempt to exchange parameters more rarely

Proposal distributions within the chains are also informed by the chain temperature.  Hotter chains should attempt large jumps in parameter space so they rapidly explore the entire prior volume, while colder chains should take smaller jumps so they efficiently accept new states in the chain.  It is natural to scale the jumps by the temperature
\begin{equation}
\delta\theta\rightarrow\delta\theta\sqrt{T_i}
\end{equation}
which is, again, motivated by the assumption that the target distribution is Gaussian.  For the hotter chains, this assumption is no longer valid, and these jumps can become large enough that they greatly exceed the prior ranges, or are never accepted if the priors are informative (i.e., have regions of high probability).  Some care needs to be taken when scaling proposals by the chain temperature to avoid quenching the free-range exploration of the hottest chains.  The hotter chains can frequently encounter the bounds of the prior volume.  This can enhance the need to carefully deal with jumps that are outside of the prior range (for instance, using jumps along eigenvectors of the Fisher matrix when the current solution is near the edge of the prior).  We have used periodic and reflecting boundary conditions on the prior volume.  Simply rejecting proposals that are out of bounds violates the detailed balance condition.

\subsection{Diagnostics}
In our experience, the most important diagnostic tool to use when testing an MCMC algorithm is to verify that the target distribution of a $T=\infty$ chain is identical to the prior distribution.  Because you can do this study without ever calculating the likelihood (often the most computationally demanding part of each iteration) this is a quick, invaluable test.

It is also important to note that burn-in times for hot and cold chains can be very different depending on the proposal distributions and the size of the prior ranges for the model parameters.  It is always a good idea, at least during the testing of the algorithm, to store each temperatures chain.

\begin{figure}[htbp]
   \centering
   \includegraphics[width=1\linewidth]{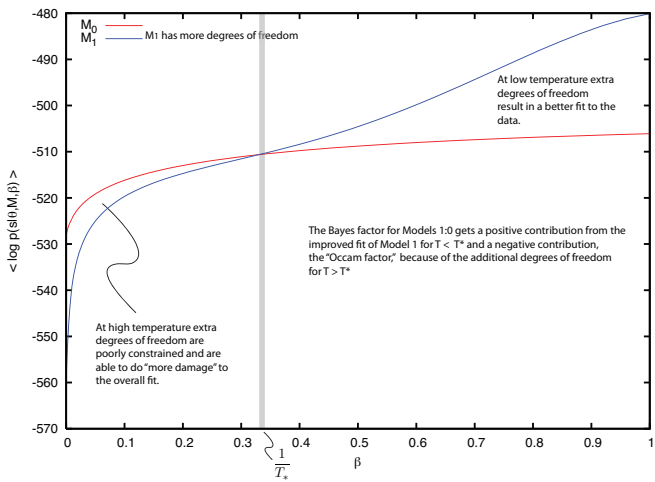} % requires the graphicx package
   \caption{An example $\langle \log \mathcal{L} \rangle$ vs. $\beta$ plot showing the location of $T_*$ and other tips to understanding how parallel tempering and thermodynamic integration work.  In this example the blue curve is for the model with a higher number of degrees of freedom, and was the favored explanation of the data.}
   \label{linear}
\end{figure}

\begin{figure}[htbp]
   \centering
  \includegraphics[width=1\linewidth]{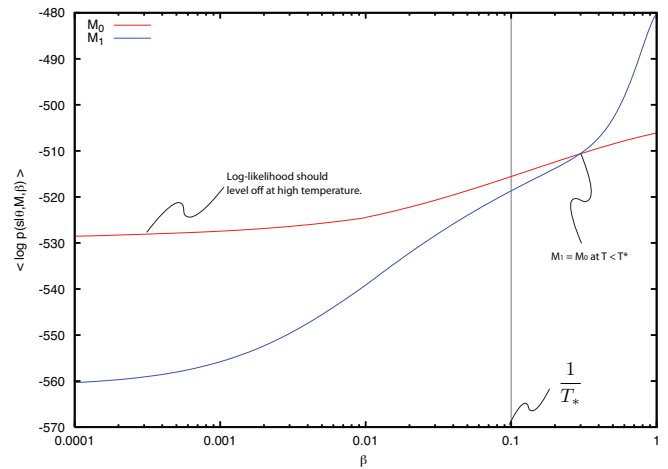} % requires the graphicx package
   \caption{The same plot as is found in figure~\ref{linear} but on a log scale in $\beta$.  The log-version of these plots is a valuable diagnostic for the behavior of the chains at high temperature.  This also shows the ``rule of thumb'' value for $T_*$ which passes the test of being lower than the point where $M_1=M_0$.}
   \label{log}
\end{figure}

Other useful figures of merit are plots of  $\langle \log p(s|\vec{\theta},\mathcal{M},\beta)\rangle$ versus $\beta$.  Figures~\ref{linear} and~\ref{log} are annotated with some of the more valuable diagnostic features.  In these examples, $M_1$ has a higher dimension than $M_0$ and was the more appropriate model.  The higher dimensional model should normally have equal or (more often) higher likelihood at low temperature than a model with fewer degrees of freedom.  Exceptions to this rule are possible if the two models are not nested (i.e., one model is not a limiting case of the other).  We have exclusively used these methods to distinguish between nested models (most often the GW detection problem) so the following discussion assumes that to be the case.

The average log-likelihood should increase smoothly and monotonically with increasing $\beta$.  A jagged $\langle p(s|\vec{\theta}) \rangle$ curve is indicative of large errors in the calculation of the average log-likelihood, typically associated with poor convergence of the chains.  Longer run times, more efficient proposal distributions, or different heating schemes can all help improve the convergence rates thus reducing the variance of the likelihood chain.

At low temperature, $M_1$ is able to achieve a better fit to the data and therefore returns a higher likelihood.  This is true even if $M_1$ is not the best description of the data due to its additional degrees of freedom.  The high temperature chains supply the ``Occam factor,'' or penalty for carrying that additional flexibility.
 
The point at which $M_0\simeq M_1$ (the ``equilibrium temperature'') indicates when the high-dimensional model chain has become sufficiently hot that the ``extra'' model parameters lose touch with whatever feature they had been fitting, and begin predominantly sampling the priors~(see figure~\ref{linear}).  At this point it is safe to switch to the larger chain spacing, however the location of $T_*$ isn't something known {\emph a priori}.  Some educated tuning/guessing is needed to determine the location of such a transition point.  We can validate the choice of $T_*$ by ensuring that it is lower than the equilibrium temperature of the chains.

The temperature range where a model switches from fitting to the data and sampling from the priors can result in a steep drop in average likelihood, especially for high dimensional models.  This large transition can be difficult to sample with chains automatically, as it can occur after the temperature spacing has been increased.  If it is poorly resolved, the exact interval between chains where change occurs is not convergent between trial runs.  Should this be the case, an additional block of finely sampled (in temperature) chains should be inserted around that region.

At high temperature, the model with more flexibility is no longer constrained by the data, and the extra degrees of freedom result in, on average, a much worse fit to the data.  

To clarify this, consider the case where the the two models under consideration are that of the detection problem (for this discussion we will assume stationary Gaussian noise and uninformative priors).  A noise-only model will not be able to accommodate a GW signal in the data, while a template will fit to the GW signal (if there is one) or to some part of the noise (if there is not).  Regardless of the data, $M_1$ will return a higher likelihood than $M_0$ for low temperature.   At high temperature, $M_1$ is sampling from the priors on the GW parameters without being significantly informed by the data because the likelihood ratio in the transition probability is suppressed by the temperature.  Therefore, high temperature $M_1$ chains are likely to include in the data a bright GW signal which can result in a large $\chi^2$ and accordingly, a low value for the likelihood.  Colloquially, $M_1$ can do a lot of ``damage'' to the residual if it is effectively unconstrained.

For thermodynamic integration, the model selection question comes down to whether or not the improvement in likelihood for the colder chains can overwhelm the Occam factor supplied by the hotter chains.  This emphasizes the importance of choosing an appropriate $T_{\rm max}$.  The average likelihood stops decreasing once the temperature is high enough that additional chains in the ladder produce samples from identical distributions (the prior).  This shows up as a floor in the likelihood versus inverse temperature plot best seen in Fig~\ref{log}.  Beyond this point, contributions to the evidence integral are proportional to $\beta$ (which is approaching zero) and can safely be neglected.

\end{document}